\documentclass[pre,twocolumn, amsmath,amssymb,aps,floatfix]{revtex4-2}
\usepackage{subcaption}
\usepackage{color,soul}
\usepackage{graphicx}
\usepackage{dcolumn}
\usepackage{bm}
\usepackage{framed,graphicx,xcolor}
\usepackage{ragged2e}
\begin{document}
\definecolor{shadecolor}{rgb}{1.0,1.0,0} 
\preprint{APS/123-QED}

\title{Geometric thermodynamics of collapse of gels}

 \author{Asif~Raza}%
\email{asifraza@iisc.ac.in}
\altaffiliation[Also at ]{Centre of Excellence on Advanced Mechanics of Materials, Indian Institute of Science.}
\affiliation{%
Computational Mechanics Laboratory, Department of Civil Engineering,
Indian Institute of Science, Bangalore, 
}
\author{Sanhita~Das}
 \email{sanhitadas@iitj.ac.in}
\affiliation{Department of Civil and Infrastructure Engineering, Indian Institute of Technology Jodhpur}

\author{Debasish~Roy}%
\email{royd@iisc.ac.in}
\altaffiliation[Also at ]{Centre of Excellence on Advanced Mechanics of Materials, Indian Institute of Science.}
\affiliation{%
Computational Mechanics Laboratory, Department of Civil Engineering,
Indian Institute of Science, Bangalore, 
}

\date{\today}

\begin{abstract}
Stimulus-induced volumetric phase transition in gels may be potentially exploited for various bio-engineering and mechanical engineering applications. Since the discovery of the phenomenon in the 1970s, extensive experimental research has helped understand the phase transition and related critical phenomena. Yet, little insight is available on the evolving microstructure. In this article, we aim at unravelling certain geometric aspects of the micromechanics underlying discontinuous phase transition in polyacrylamide gels. Towards this, we use geometric thermodynamics and a Landau-Ginzburg type free energy functional involving a squared gradient, in conjunction with Flory-Huggins theory. We specifically exploit Ruppeiner's approach of Riemannian geometry-enriched thermodynamic fluctuation theory, which was previously employed to investigate phase transitions in van der Waals fluids and black holes. The framework equips us with a scalar curvature that is typically indicative of certain aspects of the microstructure during phase transition. Since previous studies have indicated that curvature divergence relates to correlation length divergence, we infer that the gel possesses a heterogeneous microstructure during phase transition, i.e. at critical points.  Curvature also provides an insight into the universality class of phase transition and the nature of polymer-polymer interactions.

\end{abstract}

\maketitle

\section{Introduction}

Stimulus-sensitive gels, a class of polymers, find extensive applications in biomedical and mechanical engineering \cite{koetting2015stimulus, kamata2015design}. Exposed to external stimuli such as temperature, pH etc., they may undergo volumetric changes, which allows one to control their volume through stimuli manipulations. Biomedical applications where such volume control is crucial include drug delivery systems, tissue engineering, implants etc. \cite{shibayama1993volume}. 
Gels may undergo both continuous and discontinuous phase transitions \cite{tanaka1977critical, tanaka1978collapse}. The discontinuous phase transition, which is of primary interest, is similar in many ways to the phase transition in van der Waals fluids or magnetic systems \cite{shibayama1993volume, duvsek1968transition}. Both are volumetric phase transitions in systems with disordered microstructures. The swollen and shrunken phases in gels correspond respectively to the gaseous and liquid phases in van der Waals fluids. Transition in gels may however be triggered not only by temperature, but several other means such as composition of the solvent, pH and ionic changes, irradiation by light and electric fields \cite{amiya1987phase,katayama1985phase,hirokawa1984volume,hirotsu1987volume,otake1988preparation,inomata1990phase,otake1990thermal,siegel1988ph,hirokawa1984microbial,ricka1984swelling,ohmine1982salt,siegel1988ph,siegel1988ph2,osada2005conversion,tanaka1982collapse,giannetti1988molecular}. These features make such materials attractive for biomedical applications. 

Understanding the phase behavior of gels, particularly the processes of phase separation and spinodal decomposition, is crucial for optimizing their performance in various applications. Phase equilibrium based on Flory-Huggins theory offers the foundational context necessary for an understanding of phase transition \cite{flory1953principles}; however it does not account for the kinetics and the evolution of morphology with time in the separation process. To address this, the notion of gradient energy is introduced through the Landau-Ginzburg functional \cite{cahn1958free,cahn1959free}.
Landau-Ginzburg theory is often applied to understand the behavior of order parameters near critical points and track the transition among different phases  \cite{nauman1989phase,binder1987theory,binder1987dynamics,ariyapadi1990gradient,he1996phase,NAUMAN2001}. 

Phase transition and the associated critical phenomena are in need of a more nuanced treatment affording an insight into the underlying molecular interactions \cite{li1992phase}. Experiments such as dynamic light scattering, friction measurement, calorimetry \cite{hochberg1979spinodal,tanaka1977critical, hirotsu1987phase, hirotsu1987volume, li1992phase, tokita1991friction} have established the existence of a critical point and led to an understanding of the various response functions at the critical point. Several observations have come to light, viz. increase and subsequent divergence of the intensity and correlation of  scattered light, lowering and subsequent divergence of the swelling and collapsing relaxation times, and a decrease of the bulk modulus to zero \cite{li1992phase, hochberg1979spinodal}. Singularities in the specific heat, osmotic compressibility have been characterized using critical exponents \cite{li1989study} and the behaviour of the gel at the critical point is observed to conform to a 3D Ising system.  

Similar to other systems belonging to the same universality class, there are attractive and repelling forces that cause the gel to vacillate between collapsed and swollen phases. Based on the chemical composition of gels, Tanaka surmised without sufficient evidence that the interactions could be hydrophobic, electrostatic, ionic or through van der Waals forces \cite{li1992phase}. 

An alternative approach to an accounting of microstructural interactions is the thermodynamic fluctuation theory. The classical fluctuation theory however fails where fluctuations are of the order of the system size \cite{ruppeiner1995riemannian}; thus it cannot be directly exploited for phase transitions. Ruppeiner proposed a geometric approach by introducing a Riemannian structure to the thermodynamic manifold \cite{ruppeiner1995riemannian}, which eliminates the lack of covariance associated with large fluctuations. Once equipped with a Riemannian structure that naturally enforces a certain class of frame indifference, the Ricci curvature of the thermodynamic manifold may be used to gather information on microstructural interactions.

Ruppeiner's approach has been implemented to understand the microstucture of van der Waals fluids, black holes, magnetic substances and elastomers undergoing strain-induced crystallization \cite{das2022geometric, ruppeiner2010thermodynamic, ruppeiner2015thermodynamic, ruppeiner2008thermodynamic, ruppeiner2020thermodynamic, ruppeiner2021thermodynamic, kumara2023microstructure, wei2019repulsive, wei2019ruppeiner}. The sign of the scalar curvature indicates the type of interaction among the mesoscopic entities \cite{wei2019repulsive, ruppeiner2010thermodynamic}. This quantity has also been used in \cite{wei2019repulsive} to establish a similarity between phase transitions in van der Waals fluids and AdS black holes. 

In this article, we study volumetric phase transition in homogeneous polyacrylamide gels using Ruppeiner's approach. The aim is to understand the underlying interactions. We assume that no nanoscale heterogeneities are present in the gel that might render the interpretation of phase transition as a critical phenomenon redundant \cite{habicht2015critical}. We start with a brief description of Riemannian geometry, defining the relevant quantities in Sec. \ref{thermodynamicmanifoldgeneral}, followed by an analysis in Sec. \ref{Gels_Thermodynamics} of phase transition using an approach that combines the classical mean-field theory 
with Landau-Ginzburg. We then determine the curvature for the thermodynamic manifold of present interest in Sec. \ref{Gels_Curvature}. Variations of curvature with temperature and polymer volume fraction are discussed and some predictions drawn on the microstructure and the nature of interactions during phase transition. We also analyze the divergence of curvature in the vicinity of the critical point to ascertain the universality class of phase transition.  Finally, we conclude the article in Sec. \ref{Conclusion} with a summary of observations.

\section{Riemannian Thermodynamic Manifold}\label{thermodynamicmanifoldgeneral}

The thermodynamic fluctuation theory states that the probability $P$ of fluctuations of independent variables $(x_1,x_2, ... x_n)$ in a thermodynamic system is proportional to the number of microstates that is in turn related to the distance $\Delta l$ among neighbouring fluctuation states in the thermodynamic configuration space \cite{ruppeiner1995riemannian} as shown in Eq. \eqref{Fluctuationtheory}. Thus, the closer the two states, the higher the probability of fluctuation between them. 

\begin{equation}\label{Fluctuationtheory}
P(x_1, x_2, .... x_n) \propto \exp-\big({\frac{1}{2}\Delta l^2}\big)
\end{equation}
$\Delta l$ is a (positive) length measure on the thermodynamic manifold. In the transition regime, the microstructure-driven fluctuations are likely to be pronounced and attended to by incompatibilities that render the underlying vector fields non-integrable.  An accounting of such a facet in the thermodynamic manifold could be by treating it as a curved hypersurface, i.e. a Riemannian manifold, such that $\Delta l$ has a coordinate free representation. 
 \begin{equation}\begin{split}\label{Entropicmetricpsi}
(\Delta l)^2 = &-\dfrac{1}{k_{B}T}\left(\dfrac{\partial^2{\Psi}}{\partial{T}^2}\right)(\Delta T)^{2} 
\\&+ \dfrac{1}{k_{B}T}\sum_{i,j = 2}^{n}\left(\dfrac{\partial^2{\Psi}}{\partial{x_{i}}\partial{x_{j}}}\right)\Delta x_{i}\Delta x_{j}
\end{split}\end{equation}
where $\Psi$ is the free energy density of the thermodynamic system, $k_B$ the Boltzmann constant and $x_1 = T$ the temperature of the system. 

Let us consider a Riemannian manifold with metric $\mathbf{g}$ (written as a matrix $[g]=g_{ij}$) and consisting of points representing the thermodynamic states. The distance $\Delta l$ may then be written in terms of $\mathbf{g}$ as follows,

\begin{equation}\begin{split}\label{EntropicMetric}
\Delta l^2 = \sum^n_{i,j=1} g_{ij} \Delta x_{i}\Delta x_{j}
\end{split}\end{equation}
Combining Eqs. \eqref{Entropicmetricpsi} and \eqref{EntropicMetric}, $\mathbf{g}$ may be extracted in terms of $\Psi$ as follows. 

\begin{equation}\label{MetricinFreeEnergy}
    \mathbf{g} = \dfrac{1}{k_{B}T}\left[
    \begin{array}{ccccc}
    -\dfrac{\partial^2{\Psi}}{\partial{T}^2} & 0 & 0 & \cdots & 0\\
    0 & \dfrac{\partial^2{\Psi}}{\partial{x_2}^2} & \dfrac{\partial^2{\Psi}}{\partial{x_2}\partial{x_3}} & \cdots & \dfrac{\partial^2{\Psi}}{\partial{x_2}\partial{x_n}}\\
    0 & \dfrac{\partial^2{\Psi}}{\partial{x_3}\partial{x_2}} & \dfrac{\partial^2{\Psi}}{\partial{x_3}^2} & \cdots & \dfrac{\partial^2{\Psi}}{\partial{x_3}\partial{x_n}}\\
    \vdots & \vdots & \vdots & \ddots & \vdots \\
    0 & \dfrac{\partial^2{\Psi}}{\partial{x_n}\partial{x_2}} & \dfrac{\partial^2{\Psi}}{\partial{x_n}\partial{x_3}} & \cdots & \dfrac{\partial^2{\Psi}}{\partial{x_n}^2}
    \end{array}
    \right]
\end{equation}

Tangent spaces attached to different points on a Riemannian manifold are not canonically isomorphic. Thus, to transfer (or parallelly transport) quantities of interest -- vectors, to wit, from one tangent space to another, we further define an affine connection with the Christoffel symbols $\mathbf{\Gamma}_{ij}^{k}$ given by \cite{BishopGoldberg1980,Gray1997},

\begin{equation}\label{ChristoffelSymbols}
\mathbf{\Gamma}_{ij}^{k} = g^{km}\frac{1}{2}\left(\frac{\partial{g_{im}}}{\partial{x^{j}}}+\frac{\partial{g_{jm}}}{\partial{x^{i}}}-\frac{\partial{g_{ij}}}{\partial{x^{m}}}\right)
\end{equation}
where $g^{ij}$ are components of the inverse of $[g]$. $\mathbf{\Gamma}_{ij}^{k}$ may then be used to determine the components of the fourth order Riemannian curvature tensor $\Tilde{\mathbf{R}}$,

\begin{equation}\label{RiemannianCurvatureTensor}
\Tilde{R}_{ijk}^{l} = \frac{\partial}{\partial X^{i}}\mathbf{\Gamma}_{jk}^{l}-\frac{\partial}{\partial X^{j}}\mathbf{\Gamma}_{ik}^{l}+\mathbf{\Gamma}_{im}^{l}\mathbf{\Gamma}_{jk}^{m}-\mathbf{\Gamma}_{jm}^{l}\mathbf{\Gamma}_{ik}^{m}
\end{equation}
$\Tilde{\mathbf{R}}$ may be contracted to obtain the following expressions for the second order symmetric Ricci tensor $\hat{\mathbf{R}}$ and the Ricci scalar curvature $R$ respectively:

\begin{equation}\label{RicciCurvatureTensor}
\hat{R}_{ij} = \Tilde{R}_{ikj}^{k}
\end{equation}

\begin{equation}\label{RicciScalar}
R=g^{ij}\Tilde{R}^{k}_{ikj}
\end{equation}
Recall that $g^{ij}$, being components of $[g]^{-1}$, constitute the contravariant form of the metric tensor $\mathbf{g}$. 

The system of interest has two fluctuating variables  $x_1$ and $x_2$ (a temperature and a number density respectively), for which the metric $\mathbf{g}$ can be written as:
\begin{equation}\label{Metricin2D}
    \mathbf{g} = \left[
    \begin{array}{cc}
    g_{11} & 0\\
    0 & g_{22}
    \end{array}
    \right]
\end{equation}    

Consequently, in line with \cite{ruppeiner2010thermodynamic}, the Ricci scalar curvature $R$ may be determined from Eqs. \eqref{ChristoffelSymbols}, \eqref{RiemannianCurvatureTensor}, \eqref{RicciCurvatureTensor} and \eqref{RicciScalar} as :
\begin{equation}\label{RicciScalarExprsn}
    R = \dfrac{1}{\sqrt{g}}\left[\dfrac{\partial}{\partial x_1}\left( \dfrac{1}{\sqrt{g}}\dfrac{\partial g_{22}}{\partial x_1} \right) +\dfrac{\partial}{\partial x_2}\left( \dfrac{1}{\sqrt{g}}\dfrac{\partial g_{11}}{\partial x_2} \right)  \right]
\end{equation}
where $g \equiv g_{11}g_{22}$ is the determinant of $[g]$.

\section{Thermodynamics and phase diagrams of gels}\label{Gels_Thermodynamics}

\subsection{Free energy}\label{Gel_FreeEnergy}

The total free energy density $\Psi$ of a gel has a form as in \cite{tanaka1977critical} and is hypothesized to comprise of three parts:
\begin{equation}\label{freeEnergy}
\Psi = \Psi_m + \Psi_{el} + \Psi_{surr}
\end{equation}
The first term $\Psi_m$ is the free energy of mixing while the second term $\Psi_{el}$ denotes the elastic free energy due to the expansion of the network structure. The last term $\Psi_{surr}$ is the thermal free energy with respect to a reservoir at a fixed reference temperature $T_0$.

The free energy $\Psi_{m}$ of mixing for a polymer solution \cite{flory1953principles} is given by,
\begin{equation}\label{FreeEnergyMv1}
    \Psi_m = k_{B}T\left(\dfrac{\Theta}{2T}n_1v_2 + n_{1}\ln{(1-v_2)}+ n_{2}\ln{(v_2)}\right)
\end{equation}
where $n_1$ and $n_{2}$ are the numbers of solvent molecules and polymer molecules per unit reference volume respectively, $v_1$ is the volume fraction of the solvent, $v_2$ is the volume fraction of the polymer (solute) and the latter is related to $v_1$ through $v_2 = 1-v_1$. Also, $\frac{\Theta}{2T}$ is a dimensionless quantity characterizing the interaction between polymer and solvent molecules. Here, $\Theta$ is the Flory temperature at which no interactions between polymer chains and solvent molecules exist. 

We assume that $n_{2}$ which is the number of polymer molecules is equal to zero in line with  \cite{flory1953principles,tanaka1977critical,tanaka1982collapse} incorporating the fact that individual polymer molecules are absent in the network structure. Thus, the free energy of mixing $\Psi_{m}$ can be reformulated as:
\begin{equation}\label{freeEnergyMM}
    \Psi_m = k_{B}T\left(\dfrac{\Theta}{2T}n_1v_2 + n_{1}\ln{(1-v_2)}\right)
\end{equation}

Further, if $\bar{v}$ denotes the volume of an individual solvent molecule, then $v_1 = \frac{n_1 \bar{v}}{1+n_1 \bar{v}}$ and $v_2 = \frac{1}{1+n_1 \bar{v}}$. Thus, $\Psi_m$ can be expressed in terms of $n_1$ and $\bar{v}$ as follows,

\begin{equation}\label{freeEnergyMMvf}
    \Psi_m = k_{B}T\left(\dfrac{\Theta}{2T}\left(\dfrac{n_1}{1+n_1\bar{v}}\right) + n_1\ln{\left(\dfrac{n_1\bar{v}}{1+n_1\bar{v}}\right)}\right)
\end{equation}

We also assume that the gel is not completely devoid of the solvent in its reference configuration. If the volume fraction of polymer in the reference configuration is $v_{2_0}$ with the number of solvent molecules being $n_{1_0}$, we can express the elastic free energy density $\Psi_{el}$ in terms of the volumetric deformation $v_2/v_{2_0}$ as follows,

\begin{equation}\label{freeEnergyEl}
    \Psi_{el} = \dfrac{k_B T\nu_e}{2}\left(3\left(\dfrac{v_2}{v_{2_0}}\right)^{-2/3} - 3 + \ln{\left(\dfrac{v_2}{v_{2_0}}\right)} \right)
\end{equation}
where $\nu_e$ is the number of freely-moving chains between successive entanglements or cross-links. Substituting the expressions for $v_{2}$ and $v_{2_0}$ in Eq. \eqref{freeEnergyEl}, we obtain,

\begin{equation}\label{freeEnergyEl2}\begin{split}
        \Psi_{el} = \dfrac{3}{2}k_B T\dfrac{S}{\bar{v}(1+n_{1_0}\bar{v})^2}\Bigg(\left(\dfrac{1+n_{1_0}\bar{v}}{1+n_1\bar{v}}\right)^{-2/3} - 1  \\  +\ln{\left(\dfrac{1+n_{1_0}\bar{v}}{1+n_1\bar{v}}\right)^{1/3}}\Bigg)
\end{split}
\end{equation}
where $S = \nu_e\bar{v}(1+n_{1_0}\bar{v})^2$.

The thermal free energy $\Psi_{surr}$ may be obtained by successively integrating the $C_v$ of the system \cite{simo1992associative}. This is motivated from non-equilibrium thermodynamic framework such as rational thermodynamic framework which invokes the local equilibrium hypothesis. However, $C_v = - T \dfrac{\partial^2 \Psi}{\partial T^2}$ implies that $C_v = - T \dfrac{\partial^2 \Psi_{surr}}{\partial T^2}$ as
$\Psi = \Psi_{surr} + \Psi_m + \Psi_{el}$ and $\dfrac{\partial^2 \Psi_m}{\partial T^2}$ and $\dfrac{\partial^2 \Psi_{el}}{\partial T^2}$ are both identically $0$. 

\cite{li1989study} contains an expression for the specific heat capacity for temperatures lower or higher than the critical temperature for the polyacrylamide gel. It is well-established that the specific heat capacity diverges at the critical temperature \cite{hahne2005critical}, and the expression in \cite{li1989study} is consistent with the usual notion. We have

\begin{equation}\label{Cv}
    C_v(T) = A|t|^{-\alpha_{\pi}}\left[1+D|t|^{\Delta}\right] + B + C(T - T_c)
\end{equation}
where $|t|=\frac{|T-T_c|}{T_c}$ and $T_c$ is the critical temperature. Also, $\alpha_{\pi}$ is the specific heat critical exponent with a value of $-0.05$ and $\Delta$ is the correction-to-scaling exponent with a value of $0.5$. $A, B, C$ and $D$ are constants determined through curve fitting \cite{li1989study}. Their values are  $-241.581, 182, -1.193$ and $-1.115$ respectively. The thermal free energy $\Psi_{surr}$ is obtained by successively integrating $-C_{v}/T$ from \eqref{Cv} w.r.t. temperature over $T_{0}$ and $T$ : 
      
\begin{equation}\label{freeEnergySurr}
\begin{split}
    \Psi_{surr} = \Psi_c + B\left(T-T_{0}-T\ln{\dfrac{T}{T_{0}}}\right)-\dfrac{C}{2}(T-T_{0})^{2} \\
    -CT_{c}\left(T-T_{0}-T\ln{\dfrac{T}{T_{0}}}\right)
\end{split}
\end{equation}
$T_{0}$ is the reference temperature. $\Psi_{c}$ in  \eqref{freeEnergySurr} is separately expressed for $T<T_c$ and $T>T_c$. For $T<T_c$, $\Psi_c$ in Eq. \eqref{freeEnergySurr} is given by,
\begin{widetext}
 \begin{equation}\begin{split}\label{LessCriTemp}
   \Psi_c= \frac{A}{T_c}\Bigg(0.4878(T-T_{0})(T_0 - 2.9524 T_c)\left(1-\frac{T_0}{T_c}\right)^{1.05} + 0.3922 D (T-T_{0})(T_0-2.6452 T_c)\left(1-\frac{T_0}{T_c}\right)^{1.55}\\  +0.4646 T_c^2\left(\left(1-\frac{T_0}{T_c}\right)^{2.05}-\left(1-\frac{T}{T_c}\right)^{2.05}\right) + 0.253 D T_c^2\left(\left(1-\frac{T_0}{T_c}\right)^{2.55} - \left(1-\frac{T}{T_c}\right)^{2.55}\right) \\ + 0.1599 T_c^2 \left(\left(1-\frac{T_{0}}{T_c}\right)^{3.05}-\left(1-\frac{T}{T_c}\right)^{3.05}\right) \ +0.1105 D T_c^2 \left(\left(1-\frac{T_{0}}{T_c}\right)^{3.55}-\left(1-\frac{T}{T_c}\right)^{3.55}\right)\Bigg)
    \end{split}
\end{equation}   

For $T>T_c$, $\Psi_c$ assumes the form,
 \begin{equation}\begin{split}\label{MoreCriTemp}
   \Psi_c= \frac{A}{T_c}\Bigg(0.4878(T_0-T)(T_0 - 2.9524 T_c)\left(\frac{T_0}{T_c}-1\right)^{1.05} + 0.3922 D (T_0-T)(T_0-2.6452 T_c)\left(\frac{T_0}{T_c}-1\right)^{1.55}\\  +0.4646 T_c^2\left(\left(\frac{T_0}{T_c}-1\right)^{2.05}-\left(\frac{T}{T_c}-1\right)^{2.05}\right) + 0.253 D T_c^2\left(\left(\frac{T_0}{T_c}-1\right)^{2.55} - \left(\frac{T}{T_c}-1\right)^{2.55}\right) \\ - 0.1599 T_c^2 \left(\left(\frac{T_{0}}{T_c}-1\right)^{3.05}-\left(\frac{T}{T_c}-1\right)^{3.05}\right)  -0.1105 D T_c^2 \left(\left(\frac{T_{0}}{T_c}-1\right)^{3.55}-\left(\frac{T}{T_c}-1\right)^{3.55}\right)\Bigg)
    \end{split}
\end{equation}

Combining Eqs. \eqref{freeEnergyMMvf}, \eqref{freeEnergyEl2} and \eqref{freeEnergySurr}, we obtain the complete expression for $\Psi$ as,

\begin{equation}\begin{split}\label{FinalFreeEnergyN1}
    \Psi(T,n_1) = k_{B}T\left(\dfrac{\Theta}{2T}\left(\dfrac{n_1}{1+n_1\bar{v}}\right) + n_1\ln{\left(\dfrac{n_1\bar{v}}{1+n_1\bar{v}}\right)}\right) + \dfrac{3}{2}k_B T\dfrac{S}{\bar{v}(1+n_{1_0}\bar{v})^2}\left(\left(\dfrac{1+n_{1_0}\bar{v}}{1+n_1\bar{v}}\right)^{-2/3} - 1 -\ln{\left(\dfrac{1+n_{1_0}\bar{v}}{1+n_1\bar{v}}\right)^{-1/3}} \right)\\ +\Psi_c + B\left(T-T_{0}-T\ln{\dfrac{T}{T_{0}}}\right)-\dfrac{C}{2}(T-T_{0})^{2}
    -CT_{c}\left(T-T_{0}-T\ln{\dfrac{T}{T_{0}}}\right)
    \end{split}
\end{equation}
\end{widetext}

\subsection{Pressure and phase diagrams}

To characterize the phase transition and the critical phenomenon, we must obtain the relevant phase diagram. The phase diagram is a graphical representation of variations of pressure, temperature, and volume fraction. The relation among the three quantities is mathematically represented by the equation of state  for the osmotic pressure. The osmotic pressure is the thermodynamic force conjugate to the volume fraction $v_1$ 
of the solvent \cite{tanaka1977critical}.  The equation of state relating the osmotic pressure $\Pi$ with the volume fraction $v_2$ and temperature $T$ is given by,

\begin{equation}\label{Constitutiveosmoticpressure}
    \Pi = \Pi_{0} - N_A\left( \dfrac{\partial \Psi}{\partial n_1} \right)
\end{equation}
where $\Psi$ is the total free energy and $\Pi_{0}$ is the pressure of the bath in which the system has been kept.   

Using \eqref{FinalFreeEnergyN1} in \eqref{Constitutiveosmoticpressure}, we obtain the equation of state as,
\begin{widetext}
    \begin{equation}\label{Pressure}
    \Pi = \Pi_{0} -RT\left\{\dfrac{\Theta}{2T}\dfrac{1}{(1+n_{1}\bar{v})^2} + \dfrac{1}{1+n_{1}\bar{v}} + \ln{\left(\dfrac{n_1\bar{v}}{1+n_1\bar{v}}\right)} + \dfrac{S}{(1+n_{1_0}\bar{v})^3}\left(\left(\dfrac{1+n_{1_0}\bar{v}}{1+n_1\bar{v}}\right)^{1/3} - \dfrac{1}{2}\left(\dfrac{1+n_{1_0}\bar{v}}{1+n_1\bar{v}}\right)\right) \right\}
\end{equation}
\end{widetext}
where $N_A$ is Avogadro's number and $R$ the gas constant.

Eq. \eqref{Pressure} may also be written in terms of the volume fraction $v_2$ as, 

\begin{widetext}
    \begin{equation}\label{Pressurev2}
    \Pi = \Pi_{0} -RT\left\{\dfrac{\Theta}{2T}(v_2)^{2} + v_2 + \ln{\left(1-v_2\right)} + S(v_{2_0})^3\left(\left(\dfrac{v_{2}}{v_{2_0}}\right)^{1/3} - \dfrac{1}{2}\left(\dfrac{v_{2}}{v_{2_0}}\right)\right) \right\}
\end{equation}
\end{widetext}

\begin{figure}
    \centering
    \includegraphics[width=0.48\textwidth]{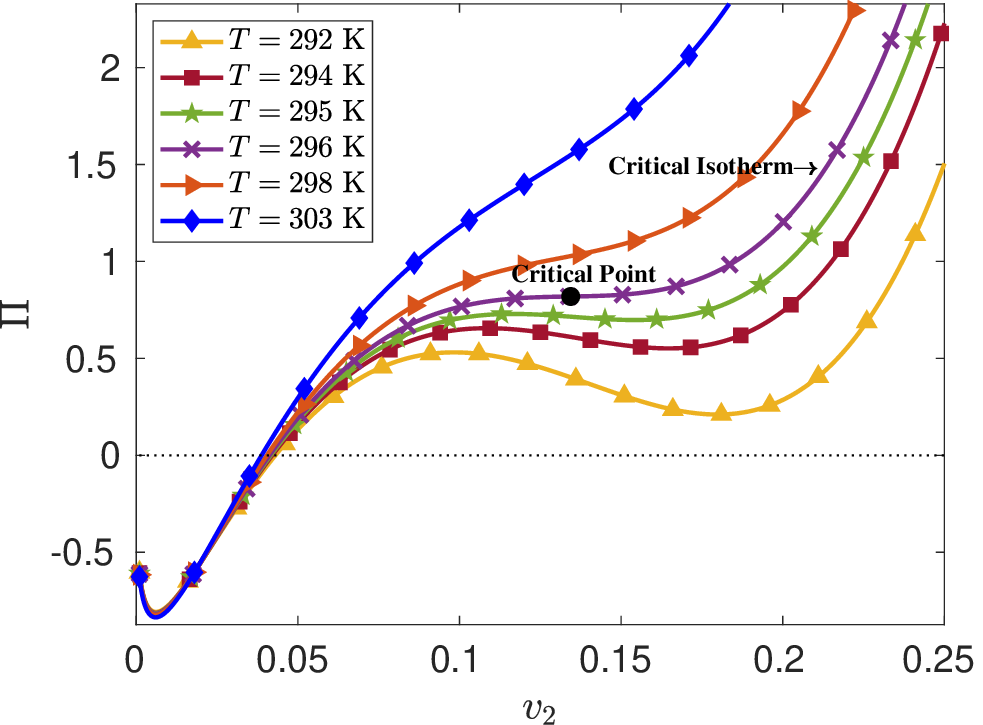}
    \caption{Plot of pressure $\Pi$ vs. volume fraction $v_2$ at various temperatures. The black dot represents the critical point and the purple line with cross markers represent the critical isotherm.}
\label{PressureModified2dPlotAllNew}
\end{figure}

Fig.~\ref{PressureModified2dPlotAllNew} shows the $\Pi-v_2$ projection of the phase diagram for a polyacrylamide gel which is obtained using Eq. \eqref{Pressurev2} (the material constants for which are available in \cite{tanaka1978collapse}): $k_B = 1.38\times10^{-23}$ J K$^{-1}$, $\Theta = 400$ K, $S = 600$, $v_{2_0} = 0.01$, $R = 8.3145 $ J K$^{-1}$ mol$^{-1}$, $\Pi_{0} = 0$ Pa. The volume of the individual solvent molecule is $\bar{v} = 3.7\times10^{-28}$ m$^3$, and this value is adopted from \cite{das2021poroviscoelasticity}.

Fig.~\ref{PressureModified2dPlotAllNew} shows the isotherms at temperatures $T = 292\text{ K}, 294\text{ K}, 295\text{ K}, 296\text{ K}, 298\text{ K and } 303\text{ K}$. In this figure, we observe all the typical features of a P-V diagram in a discontinuous phase transition. Of all the isotherms, one is the critical isotherm on which lies the critical point, a point of inflection which can be characterized by the critical osmotic pressure $\Pi_c$, the critical volume fraction $v_{2_c}$ and the critical temperature $T_c$. To determine $\Pi_c$, $v_{2_c}$ and $T_c$, we simultaneously solve $\frac{\partial \Pi}{\partial v_2} = 0$ and 
$\frac{\partial ^2 \Pi}{\partial v_2^2 } = 0$. Due to the equations being strongly nonlinear, they need to be solved numerically. For polyacrylamide gels, we use the values of constants as mentioned earlier and obtain $\Pi_c, v_{2_{c}}$ and $T_c$ as  0.82 Pa, 0.1344 and 296 K respectively, which is shown by the black dot in Fig.~\ref{PressureModified2dPlotAllNew}. Henceforth, we will use reduced pressure $\Tilde{\Pi} = \Pi/\Pi_{c}$, reduced volume fraction $\Tilde{v_{2}} = v_{2}/v_{2_{c}} $ and reduced temperature $\Tilde{T} = T/T_{c}$ to depict various phase diagrams.

Above the temperature corresponding to the critical isotherm, i.e. for $T > T_c$, distinct swollen or shrunken phases do not exist. Hence, no phase transition takes place and a single stable homogeneous "supercritical gel" phase exists. Below the critical temperature $T<T_c$, there is always a certain zone of instability, marked by negative compressibility where phase transition from a swollen to a shrunken phase takes place. Thus, the critical point serves as the boundary line. Below it, phase transition is possible; but above it, no phase transition occurs. 

Resuming our analysis on the stability of different phases, all isotherms below $T_c$ contain a zone of negative compressibility. The zone begins with a local maximum and ends with a local minimum. The locus of these extrema is the spinodal curve. Note that negative compressibility is thermodynamically infeasible and hence it is a theoretical artefact that arises from the form of the equation of state. In fact, this is also present in the P-V diagram for van der Waals fluids \cite{tanaka1978collapse,TANAKA19791404}. Physically no such state of negative compressibility exists. The gel abruptly transitions from a swollen to a shrunken phase at constant pressure. This unphysical outcome might be attributed to the absence of gradient dependent terms in the expression for free energy; such a term might indeed account for the heterogeneity occurring during phase transition.
Experimental evidence \cite{pallas1985,HIFEDA1985,Hifeda1992,CRANE1999} suggests that pressure is constant in this region even though it is in non-equilibrium. Nonetheless, the point is that, due to a thermodynamically non-equilibrium nature in this region, the conventional definition of pressure is called into question and requires further research. 

Additionally, there is an artefact present near the local minimum at the lower volume fraction value for all isotherms; it arises owing to the form of the free energy $\Psi_{el}$ and $\Psi_{m}$ as adopted from \cite{tanaka1978collapse} (see equations \eqref{freeEnergyMMvf} and \eqref{freeEnergyEl2}). 

\subsection{Modified free energy and pressure}

The free energy \eqref{FinalFreeEnergyN1} for the current system lacks convexity. Moreover, it is inadequate in modelling the spontaneous spinodal decomposition-driven phase separation \cite{binder1987dynamics,tanaka1996coarsening,NAUMAN2001} in the unstable regime. Usually, spinodal decomposition is modelled using Cahn-Hilliard or Ginzburg-Landau equation which may be derived from a functional containing a gradient-enhanced term in line with \cite{nauman1989phase,he1996phase,ariyapadi1990gradient,cahn1958free,cahn1959free}. For the volume phase transition, we consider an additional squared gradient term in the free energy with a mathematical form that is physically meaningful under nonequilibrium conditions prevalent in the unstable regime.
The additional term $\Psi_{grad}$ is given by \cite{nauman1989phase},

\begin{equation}\label{GradEnergy}
    \Psi_{grad} = \frac{\eta k_BT}{2}\left( \dfrac{1+n_1\bar{v}}{1+n_{1_0}\bar{v}} \right)(\nabla n_1)^2
\end{equation}
The multiplier $\ \frac {\eta k_BT}{2}\left( \dfrac{1+n_1\bar{v}}{1+n_{1_0}\bar{v}} \right)$ is motivated by the constant osmotic pressure in the unstable regime. $n_1$ and $\tilde{T}$ dependence of $\eta$ in the multiplier is fitted to the magnitude of the osmotic pressure at the start of the unstable regime for each isotherm in the phase diagram (see figure \ref{PressureModified2dPlotAllNew}). $\eta$ for polyacrylamide is given by $5.8484 \times 10^{4}\Tilde{T}^2 - 12.3426 \times 10^{4} \Tilde{T}+6.4762 \times 10^{4}$ and it will change for a gel with a different composition.

\begin{figure}
    \centering
    \includegraphics[width=0.48\textwidth]{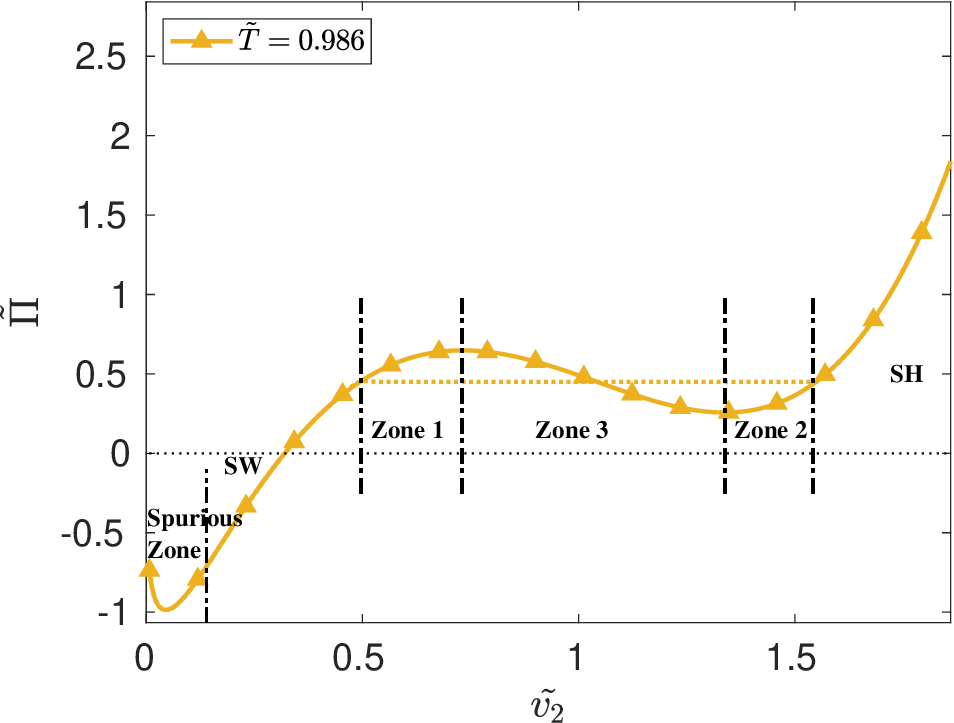}
    \caption{$\Tilde{\Pi}$-$\Tilde{v_2}$ plot for $\Tilde{T} = 0.986$ with demarcation of Zones 1, 2 and 3.}
\label{Params}
\end{figure}

The modified free energy $\Psi_{mod}$ comprises the following components - $\Psi$ given by \eqref{FinalFreeEnergyN1}, the free energy of the individual swollen and shrunken phases and $\Psi_{grad}$, the gradient-dependent energy of the mixture of phases. 
In what follows, $\kappa$ is a scaling factor varying between 0 and 1 that prioritises $\Psi_{grad}$ in the unstable regime and $\Psi$ in the stable regimes. 

\begin{equation}\label{ModEnergy}
    \Psi_{mod} = (1-\kappa)\Psi + \kappa \Psi_{grad}
\end{equation}
$\kappa$ is piecewise-defined with components $\kappa_1$, $\kappa_2$ and $\kappa_3$, 

\begin{equation}\label{kappa}
    \kappa = \kappa_{1} + \kappa_{2} + \kappa_{3}
\end{equation}
$\kappa_1$, $\kappa_2$ and $\kappa_3$ are functions dependent on $\tilde{\Pi}$ and its derivatives with non-zero values in the zones -- Zone 1, Zone 2 and Zone 3; see Fig.\ref{Params}. 

For example, $\kappa_3$ which has a non zero value in Zone 3 where $\Tilde{\Pi}$ and $-\partial \Tilde{\Pi}/\partial \Tilde{v_{2}}$ are both greater than 0 may be given by,

\begin{equation}\label{kappa3}
    \kappa_{3} = H\left(\Tilde{\Pi}\right)H\left(-\dfrac{\partial \Tilde{\Pi}}{\partial \Tilde{v_{2}}}\right)
\end{equation}
The zone boundaries have been imposed in the respective functions using the Heaviside step function denoted by $H$.
Similarly, $\kappa_1$  and $\kappa_2$ with local supports in Zone 1 and Zone 2 respectively, may be defined as

\begin{equation}\label{kappa1}
\begin{split}
    \kappa_{1} =  H\left(\exp{\left\{ -\dfrac{1}{l_{1}}\left(\dfrac{\partial \Tilde{\Pi}}{\partial \Tilde{v_{2}}} \right)^2 \right\}}\right)
    H\left(-\tanh{\left(\dfrac{\partial^2 \Tilde{\Pi}}{\partial \Tilde{v_{2}}^2} \right)}\right)
    \\
    \times(1-\kappa_{3})
\end{split}
\end{equation}

\begin{equation}\label{kappa2}
\begin{split}
    \kappa_{2} =  H\left(\exp{\left\{ -\dfrac{1}{l_{2}}\left(\dfrac{\partial \Tilde{\Pi}}{\partial \Tilde{v_{2}}} \right)^2 \right\}}\right) H\left(\tanh{\left(\dfrac{\partial^2 \Tilde{\Pi}}{\partial \Tilde{v_{2}}^2} \right)} - \epsilon\right)
    \\
    \times(1-\kappa_{3})
\end{split}
\end{equation}

For our calculations, we use $\nabla n_1 = 10^{12}$.
$l_{1}$, a temperature-dependent function, has the expression $l_{1} = -6.2563 \times 10^{5} \Tilde{T}^3 + 1.8697 \times 10^{6} \Tilde{T}^2 - 1.8625 \times 10^{6} \Tilde{T} + 6.1847 \times 10^{5}$ that approximates the width of Zone 1 with respect to the local maxima.
Similarly, $l_{2} = -5.9848 \times 10^{5} \Tilde{T}^3 + 1.7887 \times 10^{6} \Tilde{T}^2 - 1.7819 \times 10^{6} \Tilde{T} + 5.9175 \times 10^{5}$ is another temperature dependent function that approximates the width of Zone 2. Introducing $\epsilon = 0.1$ in the expression eliminates the consideration of the spurious zone, which contains the local minimum arising as a numerical artefact. 
In a future attempt, we may relate the functions $l_1$ and $l_2$ to the spinodal curve, so as to modify the non-convex free energy expressions in the unstable regime of first-order phase transitions.

\begin{figure}
    \centering
    \includegraphics[width=0.48\textwidth]{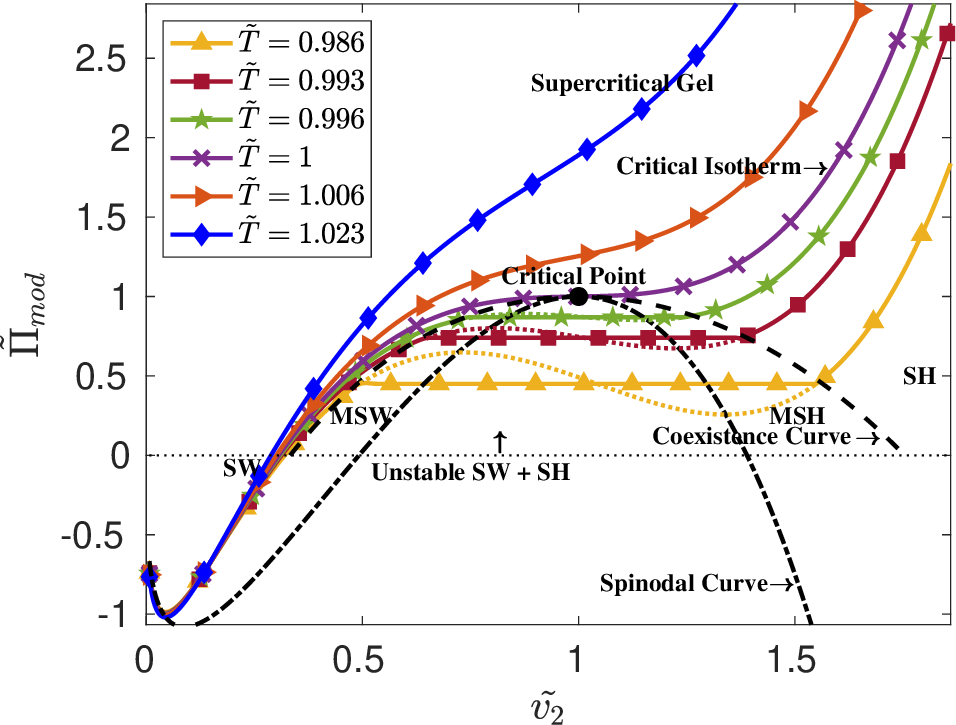}
    \caption{$\Tilde{\Pi}_{mod}$-$\Tilde{v_2}$ phase diagram denoting the various phases and containing the various isotherms. The black dot denotes the critical point and the black dashed curve represents the coexistence curve. The spinodal curve is represented by the black dash-dotted curve.}
\label{PressureModified2dPlotAll}
\end{figure}

Using $\Psi_{mod}$, we obtain $\Pi_{mod}$ from the following constitutive equation
\begin{equation}\label{PressureMod}
    \Pi_{mod} = \Pi_{0} - N_A\left( \dfrac{\partial \Psi_{mod}}{\partial n_1} \right)
\end{equation}
With $\Pi_{mod}$, the isotherms are determined and shown as bold lines in the $\tilde{\Pi}_{mod}$ vs $\tilde{v}_2$ phase diagram as shown in Fig.~\ref{PressureModified2dPlotAll}. The plots have been obtained using dimensionless pressure ($\tilde{\Pi}_{mod} = \Pi_{mod}/\Pi_c $), volume fraction ($\tilde{v}_2 = v_2/v_{2c} $) and temperature ($\tilde{T} = T/T_{c} $).  Note that in the unstable regime, pressure is constant in line with phase transition experiments on gels \cite{pallas1985,HIFEDA1985,Hifeda1992,CRANE1999}. The phase diagram also shows the coexistence curve and the spinodal curve. Between the coexistence (black dashed) and spinodal (black dotted-dashed) curves, the metastable phases - predominantly swollen metastable (MSW) and predominantly shrunken metastable -- phases exist. Also shown are the homogeneously swollen (SW) and homogeneously shrunken (SH) phases at low and high volume fractions respectively. The coexistence and spinodal curves meet at the critical point, as indicated in the figure.

The isotherms obtained from the modified free energy yield the same pressures as those obtained by the Maxwell tie line construction \cite{clerk1875dynamical}. Maxwell construction is a graphical method that provides correction to the isotherms in the P-V diagram of first-order phase transitions, in cases where the free energy expression violates convexity requirements. The tie-line is drawn subtending equal areas between the local maximum and minimum and the tie-line. At any point on the tie-line, both swollen and shrunken phases co-exist in a phase-separated microstructure, and the relative volume fraction of each phase may be determined using the lever-arm rule. Our prescribed method of working with a modified free energy could thus have implications beyond fixing the convexity requirements in the sense that the modified free energy can be directly used to obtain physically relevant isotherms without a rigorous graphical process.
However, the modified free energy is not strongly convex in the sense that the free energy is  linear over the phase transition regime. In other words, in the language of large deviation theory, this would mean that the corresponding rate function is multimodal. In our future articles we plan to explore additional large deviation theory based approaches to strongly convex free energy forms.

\begin{figure}
  \centering
\includegraphics[width=0.48\textwidth]{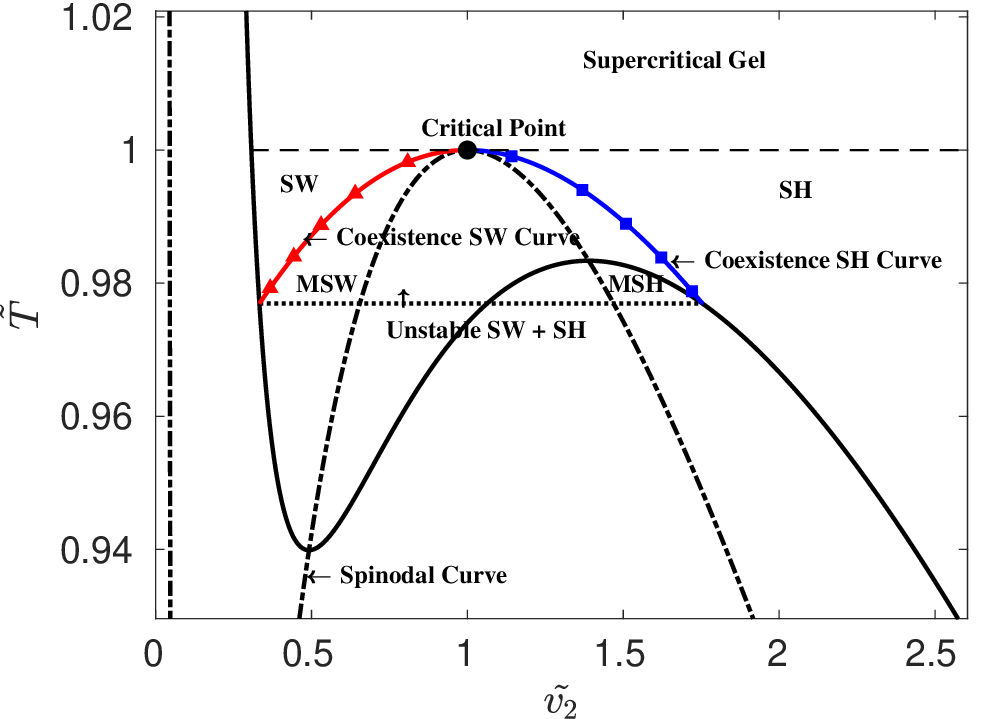}
\caption{$\Tilde{T}-\Tilde{v}_2$ phase diagram  denoting the various phases. The red solid curve with triangular markers and the blue solid curve with square markers represent the swollen and shrunken limbs of the coexistence curve respectively. The black dot denotes the critical point and the black solid curve represents the isobar at $\Tilde{\Pi}_{mod} = 0$. The dotted horizontal line at $\Tilde{T} = 0.976$ corresponds to the phase transition temperature and the black dash-dotted curve represents the spinodal curve.}
\label{SwellingCurve}
\end{figure}

\begin{figure}
  \centering
\includegraphics[width=0.48\textwidth]{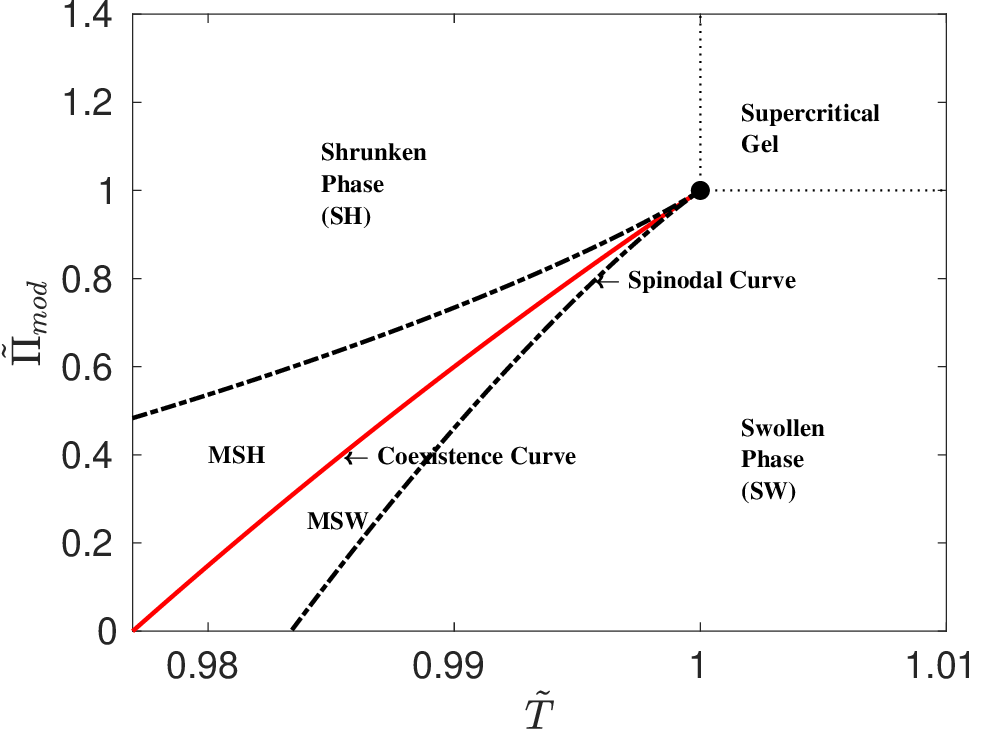}
\caption{$\Tilde{\Pi}_{mod}-\Tilde{T}$ phase diagram denoting the various phases. The red solid curve represents the coexistence curve. The black dash-dotted curve is the spinodal curve. The black dot represents the critical point.}
\label{PressureVsTemp}
\end{figure}

Fig.~\ref{SwellingCurve} shows the $\Tilde{T}-\Tilde{v_2}$ projection of the phase diagram. The solid black curve denotes the isobar $\Pi_{mod}/\Pi_c = 0$. Coexistence and spinodal curves may be seen in  $\Tilde{\Pi}_{mod} - \Tilde{v}_2$ projection as well. The swollen limb of the coexistence curve, where the predominately swollen phase exists, is represented by the solid red curve with triangular markers, while the shrunken limb of the coexistence curve is represented by the solid blue curve  with square markers in Fig.~\ref{SwellingCurve}. 

Finally, the $\Tilde{\Pi}_{mod}-\Tilde{T}$ projection is shown in Fig. \ref{PressureVsTemp}. In this figure, the entire coexistence or the phase transition region described in the earlier diagrams (Fig. \ref{PressureModified2dPlotAll} and Fig. \ref{SwellingCurve}) appears as the solid red line. The shrunken and swollen phases outside the phase transition region are also shown in the figure. Also shown are the critical point and the supercritical gel region. The spinodal curve has also been depicted by the black dash-dotted curve in Fig.~\ref{SwellingCurve} and Fig.~\ref{PressureVsTemp}.

\section{Thermodynamic geometry of gels}\label{Gels_Curvature}

The phase diagrams afforded a certain measure of understanding of phase transition in our system. Now, using Ruppeiner's approach, we attempt to track the evolving microstructure throughout this process. Indeed, the nonlinear nature of the manifold on which the microstructural evolution happens should ideally rule out the emergence of unstable regions. Writing  Eq. \eqref{Entropicmetricpsi} in terms of the fluctuating variables $T$ and $n_1$, we obtain the following expression for the squared length of an incremental line element,

\begin{equation}\label{LineElement}
(\Delta l)^2 = -\dfrac{1}{k_{B}T}\left(\dfrac{\partial^2{\Psi_{mod}}}{\partial{T}^2}\right)(\Delta T)^{2} 
+ \dfrac{1}{k_{B}T}\left(\dfrac{\partial^2{\Psi_{mod}}}{\partial{n_1}^2}\right)(\Delta n_1)^{2} 
\end{equation}
where $\Psi_{mod}$ is the modified free energy as defined in the Eq.\eqref{ModEnergy} with $\Psi_{mod} = (1-\kappa)\Psi + \kappa \Psi_{grad}$.  
The (matrix form of the) metric tensor in Eq. \eqref{MetricinFreeEnergy} is given by

\begin{equation}\begin{split}\label{Metric}
    \mathbf{g} &= \dfrac{1}{k_{B}T}\left[
    \begin{array}{cc}
    -\dfrac{\partial^2{\Psi_{mod}}}{\partial{T}^2} & 0 \\
    0 & \dfrac{\partial^2{\Psi_{mod}}}{\partial{n_1}^2} \\
    \end{array}
    \right]
    \end{split}
\end{equation}
\begin{equation}\begin{split}\label{Metric2}
    \mathbf{g} &= \dfrac{1-\kappa}{k_{B}T}\left[
    \begin{array}{cc}
    -\dfrac{\partial^2{\Psi}}{\partial{T}^2} & 0 \\
    0 & \dfrac{\partial^2{\Psi}}{\partial{n_1}^2} \\
    \end{array}
    \right] + \dfrac{\kappa}{k_{B}T}\left[
    \begin{array}{cc}
    -\dfrac{\partial^2{\Psi_{grad}}}{\partial{T}^2} & 0 \\
    0 & \dfrac{\partial^2{\Psi_{grad}}}{\partial{n_1}^2} \\
    \end{array}
    \right]
    \end{split}
\end{equation}
Here, $\kappa$ itself is a complex function of higher-order derivatives or product of derivatives of the free energy, making the contribution of its derivative to the components of the metric tensor negligible. Also, $\dfrac{\partial^2{\Psi_{grad}}}{\partial{n_1}^2}=0$ as $\Psi_{grad}$ is a linear function of $n_1$ by construction (see equation \eqref{GradEnergy}). 

In terms of $T$ and $n_1$, the Ricci scalar curvature $R$ may be determined using Eq. \eqref{RicciScalarExprsn},
\begin{equation}\label{RicciScalarExprsn2}
    R = \dfrac{1}{\sqrt{g}}\left[\dfrac{\partial}{\partial T}\left( \dfrac{1}{\sqrt{g}}\dfrac{\partial g_{22}}{\partial T} \right) +\dfrac{\partial}{\partial n_1}\left( \dfrac{1}{\sqrt{g}}\dfrac{\partial g_{11}}{\partial n_1} \right)  \right]
\end{equation}
where $g = -\dfrac{1}{(k_{B}T)^2}\dfrac{\partial^2{\Psi_{mod}}}{\partial{T}^2}\dfrac{\partial^2{\Psi_{mod}}}{\partial{n_1}^2}$ is the determinant of the metric tensor. 

Using the free energy density (Eq. \eqref{ModEnergy}) in Eq. \eqref{RicciScalarExprsn2}, the expression for $R$ outside the zone of compressibilty i.e. $\kappa = 0$ is obtained as,
\begin{equation}
    R = \dfrac{p}{q}
\end{equation}
where,
\begin{widetext}
\begin{equation}\label{numerator}
    p = - k_B\Theta n_1 \bar{v}s
    \begin{bmatrix}
    C_v\Tilde{T}(1+n_{1_0}\bar{v})(r+s) +C_v \left\{ \Tilde{T} s + \Tilde{T}\bar{v}n_1 r + \Tilde{T} r + \bar{v}n_1 s \left(\Tilde{T}-\dfrac{\Theta}{T_c}\right) \right\}\\ + \Tilde{T}\left\{ \Tilde{T} r  + \Tilde{T} \bar{v} n_1 r + \Tilde{T} s    + \bar{v}n_1 s\left(\Tilde{T}-\dfrac{\Theta}{T_c}\right)\right\}\left\{CT_c+\dfrac{0.05A(\Tilde{T} -1)}{|\Tilde{T} -1|^{1.95}}+\dfrac{0.55AD(\Tilde{T} -1)}{|\Tilde{T} -1|^{1.45}}\right\}
    \end{bmatrix}
\end{equation}
\begin{equation}\label{denominator}
    q = 2C_v^2
    \begin{bmatrix}
   \Tilde{T} r  + \Tilde{T} \bar{v} n_1 r + \Tilde{T} s    + \bar{v}n_1 s\left(\Tilde{T}-\dfrac{\Theta}{T_c}\right)
    \end{bmatrix}^2
\end{equation}
\end{widetext}
\begin{itemize}
    \item[] $r = S\bar{v}n_1\left\{ -2 + 3\left(\dfrac{1+n_{1_0}\bar{v}}{1+n_{1}\bar{v}} \right)^{2/3} \right\}$,
    \item[] $s = 6 (1+n_{1_0}\bar{v})^2 \left(\dfrac{1+n_{1_0}\bar{v}}{1+n_{1}\bar{v}} \right)^{2/3}$.
\end{itemize} 
where $S$ is the constant defined in \eqref{freeEnergyEl2}. 

\begin{figure}
  \centering
\includegraphics[width=0.48\textwidth]{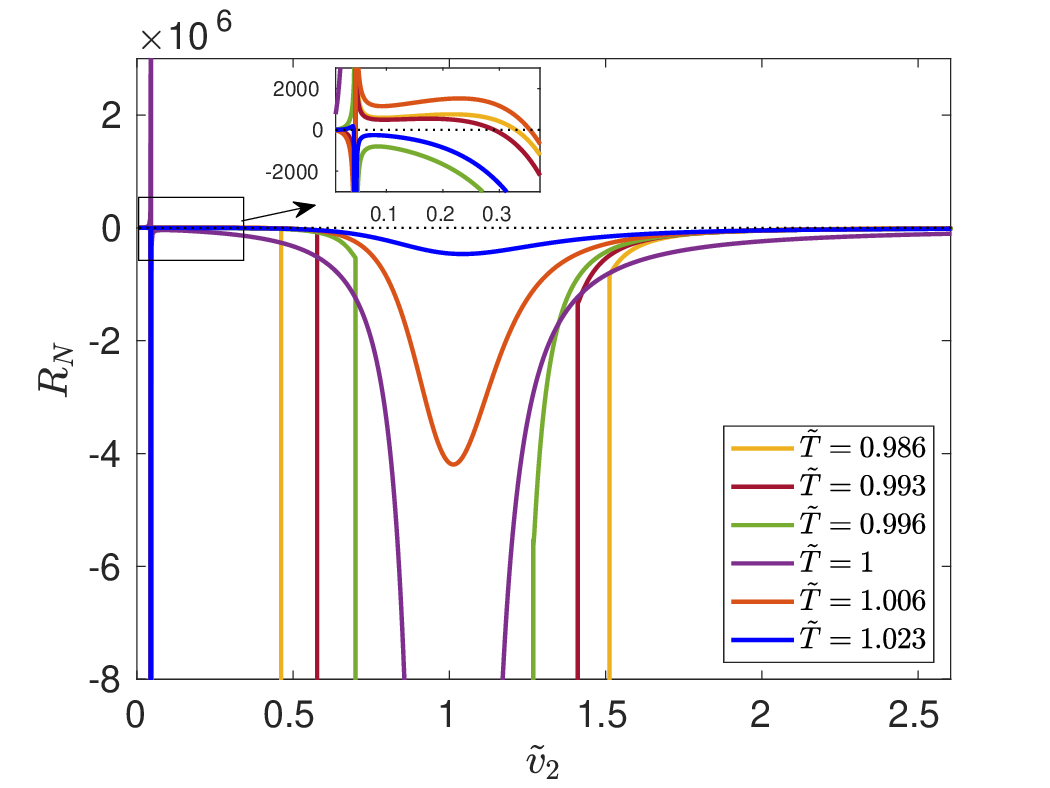}
\caption{Plots of normalized scalar curvature $R_{N}$ with volume fraction $\Tilde{v_2}$ at various temperatures}
\label{Curvature2dPlot}
\end{figure}

\begin{figure}
  \centering
\includegraphics[width=0.48\textwidth]{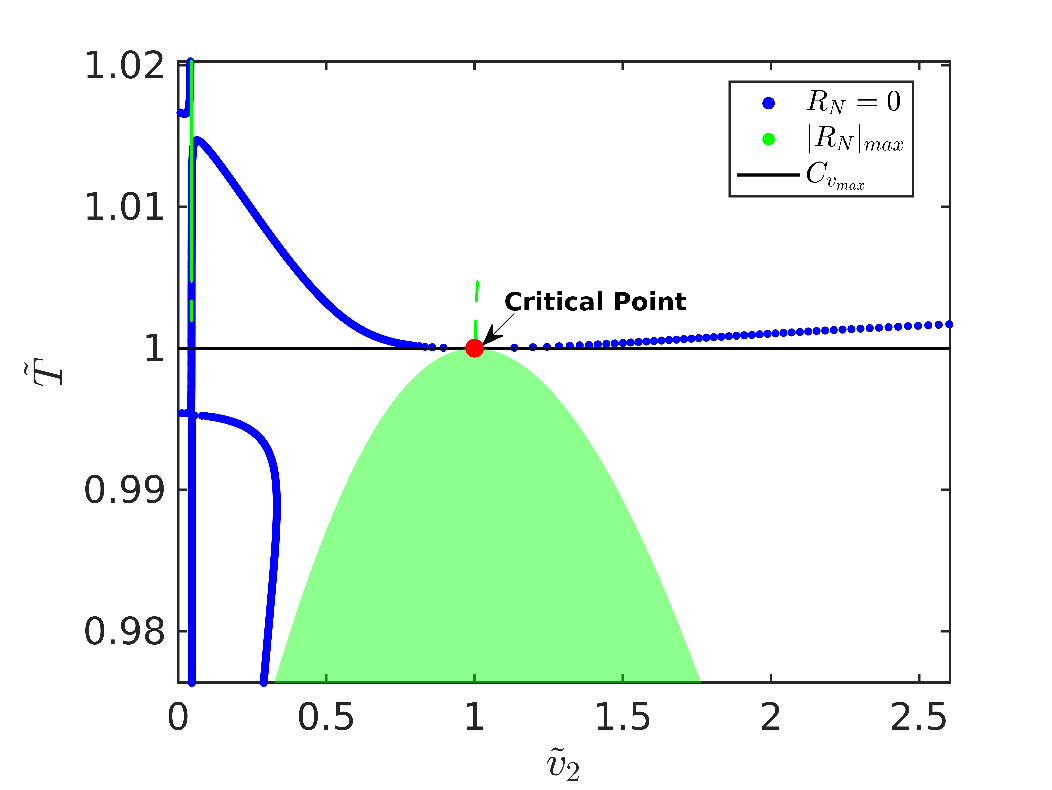}
\caption{$\Tilde{T}-\Tilde{v_{2}}$ phase diagram depicting the points where $R_{N} = 0$ and $|R_{N}|$ is maximum for a particular temperature $\Tilde{T}$. Red circular marker represents the critical point and black solid curve represents the $C_{v_{max}}$.}
\label{ZeroR}
\end{figure}

 In our geometric setting, the modification to the free energy is significant as the metric is rendered zero instead of negative in the unstable regime by enforcing $\Psi_{mod} = \Psi_{grad}$. Due to this construction, the scalar curvature $R$ becomes $-\infty$ in the transition region. A large deviation theory based construction to a strongly convex free energy will perhaps yield a non-infinite curvature in the transition regime which needs to be explored further.

We define the normalized scalar curvature $R_N = \dfrac{R}{k_B}$ in line with \cite{wei2019repulsive} and observe its variation with reduced volume $\Tilde{v}_2$ for reduced temperatures $\Tilde{T} = 0.986, 0.993, 0.996, 1, 1.006$ and $1.023$; the results are shown in Fig. \ref{Curvature2dPlot}. The curvature assumes small to moderate values over most of the parametric space except in the vicinity of the spinodal curve and the coexistence region, where it diverges. For $\Tilde{T} = 1$, divergence of $R_N$ is observed only at the critical point $v_{2_c} = 1$. For $\Tilde{T} > 1$, divergence of $R_N$ is not observed. Additionally, $R_N$ is observed to be positive for a certain region of the phase space, but not consistently so for all temperatures. This appears to be a numerical artefact, which nevertheless requires a more careful assessment in the future. Due to the lack of a consistent trend, this change of sign cannot be attributed to the change in the nature of inter-molecular interactions as has been classically done in the case of all Ruppeiner's theories.

In Fig.~\ref{ZeroR}, the $\Tilde{T}-\Tilde{v_{2}}$ projection is shown with the curve $R_{N} = 0$, $C_{v_{max}}$, the set of points corresponding to $|R_{N}|_{max}$, and the critical point. The critical point has been shown with a red solid circular marker. The black solid line represents $C_{v_{max}}$ which is a straight line parallel to the volume fraction axis passing through the critical temperature $\Tilde{T}=1$ as $C_{v}$ is solely dependent on the temperature $T$, (see Eq.~\eqref{Cv}).
$R_{N} = 0$ is shown with blue markers. Note that for every isotherm $\Tilde{T}$ except near the critical regime, there are two points where $R_{N}$ is zero. Of these, one is attributed to the presence of artefact of local minimum in the lower volume fraction that arises due to the free energy form $\Psi_{el}$ \& $\Psi_{m}$ adopted from \cite{tanaka1978collapse}.

$|R_{N}|_{max}$ is shown in green (light gray) in Fig.~\ref{ZeroR}. For $\tilde{T}\le 1$, the transition region is where $|R_{N}|_{max}$ persists as shown in the figure. For $\Tilde{T}>1$, $|R_{N}|_{max}$ shows an erratic behavior in the region of lower volume fraction values which is again attributed to the presence of artefact of local minimum in the lower volume fraction. Had this not been present, $|R_{N}|_{max}$ would have been a continuous line emanating from the critical point for $\Tilde{T}>1$.

\begin{figure}
     \centering
     \begin{subfigure}[b]{0.45\textwidth}\label{SW}
         \centering
         \includegraphics[width=0.4\textwidth]{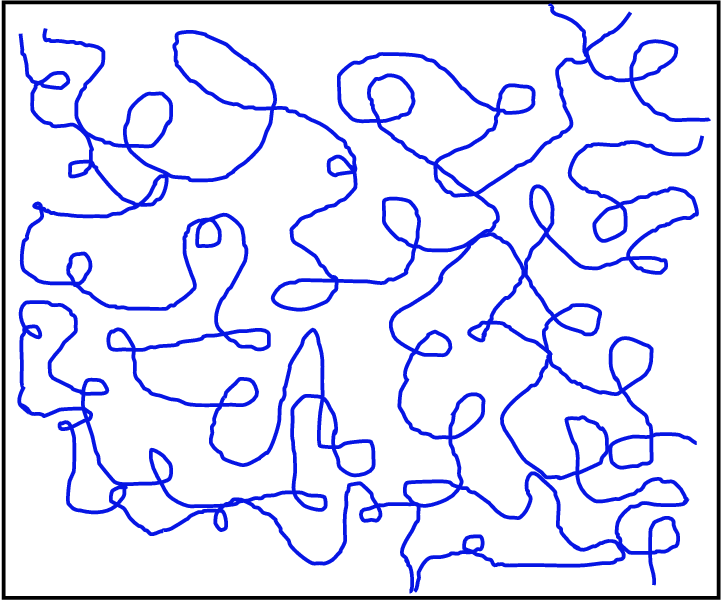}
         \caption{Swollen phase (\textbf{SW})}
     \end{subfigure}
          \begin{subfigure}[b]{0.45\textwidth}\label{SH}
         \centering
         \includegraphics[width=0.4\textwidth]{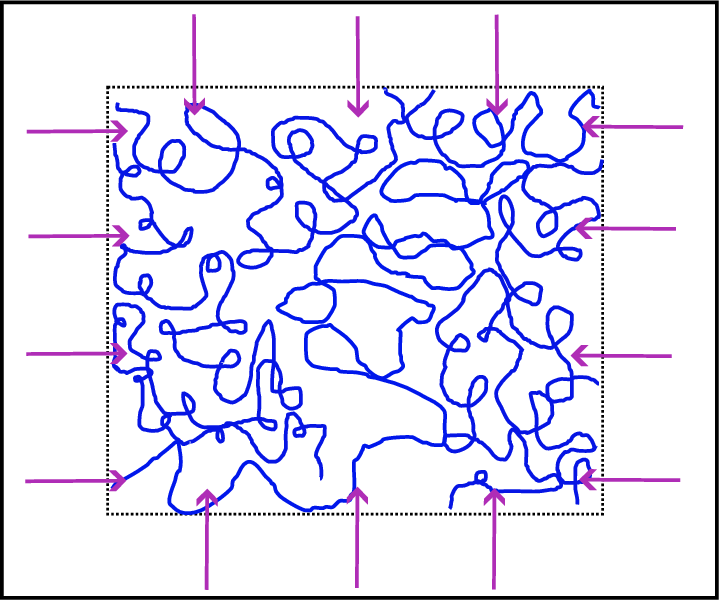}
         \caption{Shrunken phase (\textbf{SH})}
     \end{subfigure} 
     \begin{subfigure}[b]{0.45\textwidth}\label{SWSH}
         \centering
         \includegraphics[width=0.4\textwidth]{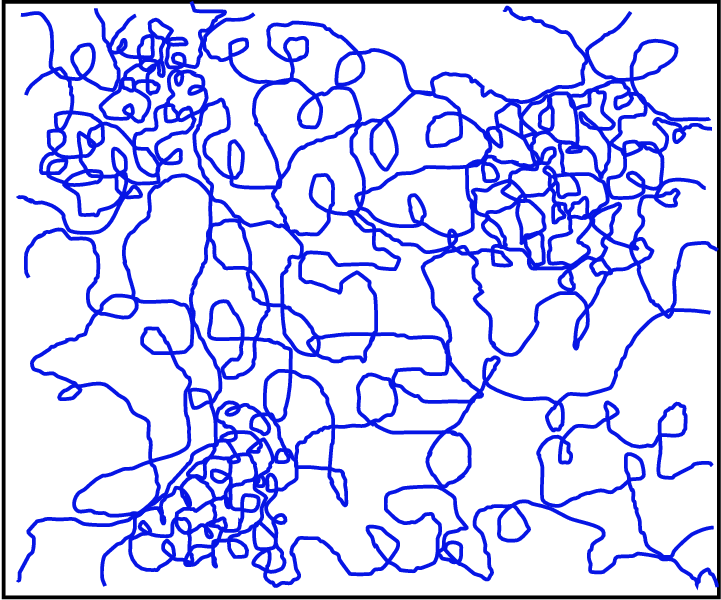}
         \caption{Heterogeneous state with both swollen phase and shrunken phase (\textbf{Unstable SW + SH})}
     \end{subfigure} 
     \begin{subfigure}[b]{0.45\textwidth}\label{FigCritical}
         \centering
         \includegraphics[width=0.4\textwidth]{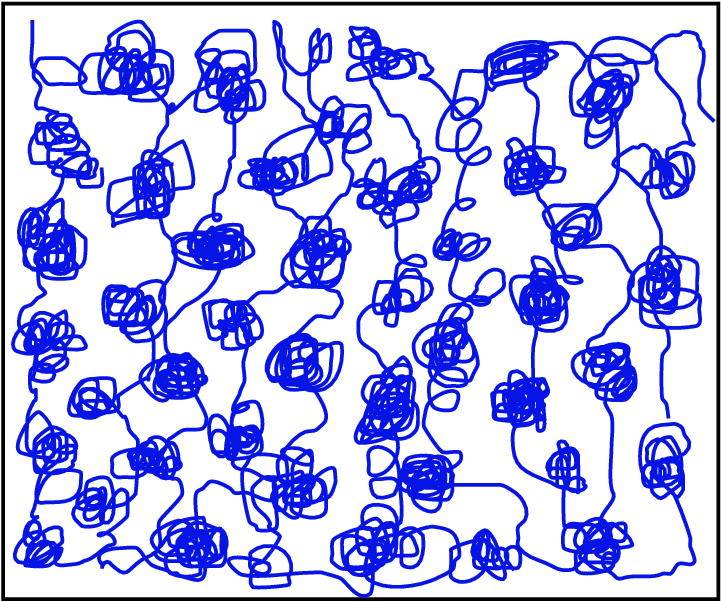}
         \caption{Gel phase at critical point}
     \end{subfigure} 
\caption{{Schematic representation of the microstructure of different phases of the gel}}
\label{Microstructure}
\end{figure}

\begin{figure}
     \centering
         \includegraphics[width=0.48\textwidth]{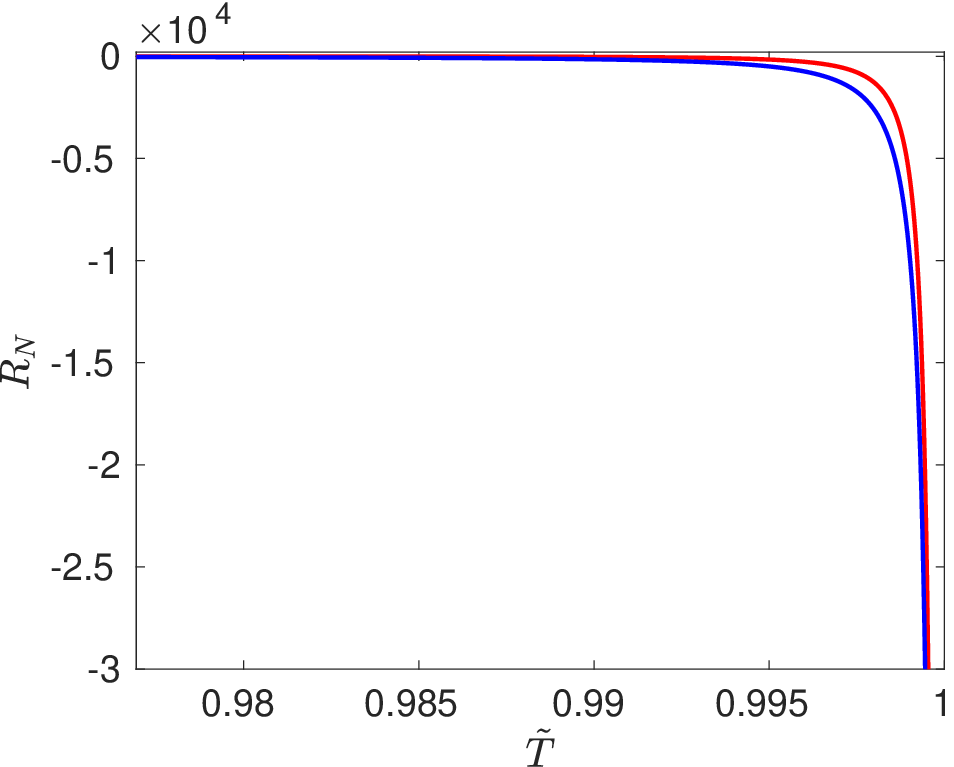}
\caption{Plot showing the behavior of normalized scalar curvature $R_{N}$ along the swollen limb (top red line) and shrunken limb (bottom blue line) of the coexistence curve.}
\label{CurCoexistence}
\end{figure}

\begin{figure*}
     \centering
     \begin{tabular}{cc}
     \begin{subfigure}[b]{0.45\textwidth}
         \centering
         \includegraphics[width=\textwidth]{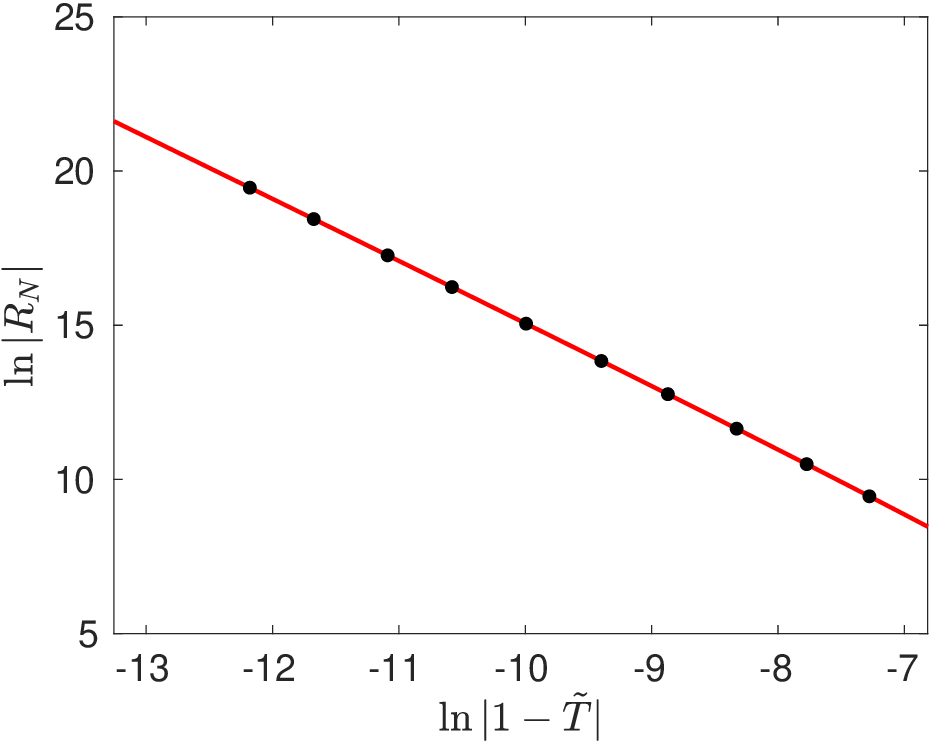}
         \caption{}
     \end{subfigure}
     \begin{subfigure}[b]{0.45\textwidth}
         \centering
         \includegraphics[width=\textwidth]{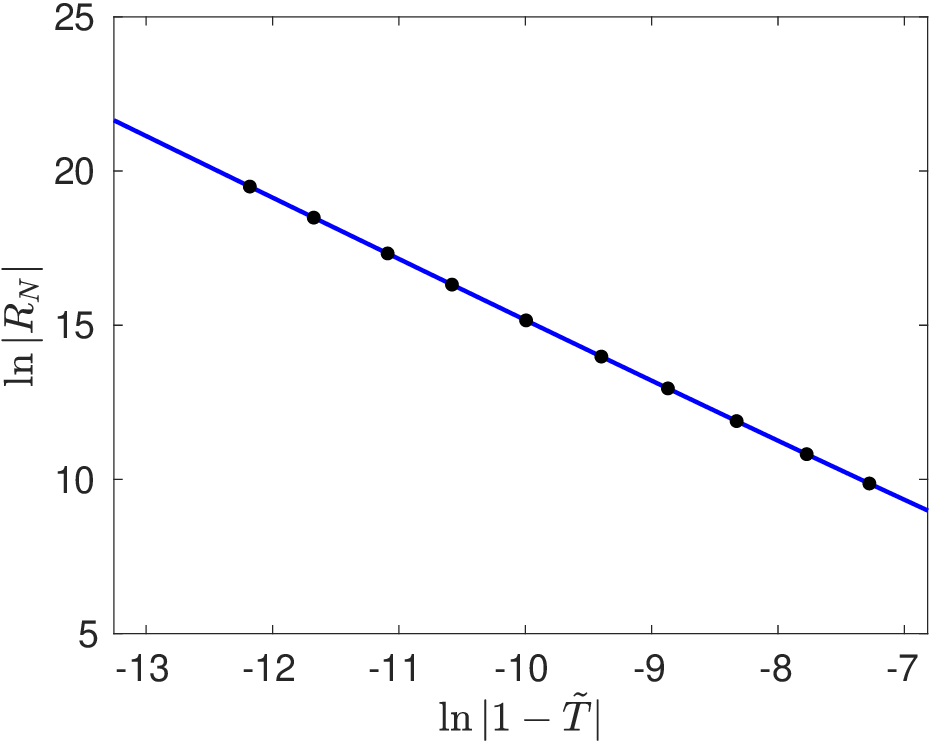}
         \caption{} 
     \end{subfigure}
     \end{tabular}
     
\caption{Plot of $\ln|R_{N}|$ vs $\ln|1-\Tilde{T}|$. Black markers denote the numerically obtained data and the solid line denotes the fitted curve. (a) The coexistence swollen phase curve (red line)---the slope is -2.00902. (b) The coexistence shrunken phase curve (blue line)---the slope is -1.99123.}
\label{CurCoexistenceSWSHLog}
\end{figure*}
In the context of black holes, divergence of $R_N$ is speculated to have some relation with divergence of the correlation length $\xi$  \cite{wei2019repulsive}. We believe that the same holds true for volumetric phase transition and divergence in curvature is directly related to that in the fluctuations of the order parameter $\Tilde{v}_2$. Thus, density fluctuations are divergent in the coexistence region and at the critical point. Within the coexistence region, density fluctuations manifest as phase segregation at the macroscopic scale, attributed to spinodal decomposition within the spinodal or unstable region and, nucleation and growth in the metastable region. At the critical point,  fluctuations cause segregation at the microscopic scale which manifests as critical opalescence in light scattering experiments. Photon correlation spectroscopic studies on polyacrylamide gels \cite{tanaka1977critical} have established the existence of a phase-segregated microstructure on the coexistence curve, which confirms our understanding. In Fig. \ref{Microstructure}, we have shown some schematic representations of the microstructure in swollen, shrunken, critical and unstable regimes based on our interpretation of curvature and aided by the work in \cite{tanaka1977critical}. The gel appears homogeneous in the swollen and shrunken phases and heterogeneous in the unstable and critical regimes. Based on what is known of the Van der Waals fluid system \cite{wei2019repulsive}, we may expect a homogeneous microstructure in the supercritical regime but with large microscale density fluctuations; see Fig.\ref{Curvature2dPlot}. But owing to a lack of experimental data pertaining to $C_v$ at higher temperatures, further insight at this stage is not available.

Curvature has also been exploited to reveal the nature of interactions among micro-constituents during different stages of phase transition in van der Waals fluids and black holes. For example, the sign of curvature in the case of van der Waals fluid  \cite{ruppeiner2020thermodynamic} indicates whether the predominant force between fluid molecules is attractive or repulsive for a particular temperature or volume fraction. We examine the variation in the sign of curvature along the coexistence curve on lines similar to \cite{wei2019repulsive}. The curvature remains negative for all temperatures leading to the conclusion that no radical change occurs in the nature of interaction with change in volume fraction $v_2$ over the entire range of temperatures considered. 


Finally, we would like to examine the critical behaviour of $R_N$ near the critical point to establish the universality class of phase transition further. We start by examining the critical exponent with which the curvature diverges. To evaluate the exponent, we resort to a numerical technique described in \cite{wei2019ruppeiner}. First we assume a critical-temperature dependent functional form for $R_N$ in the vicinity of the critical point and fit the curve to discrete points lying on the swollen and shrunken limbs of the coexistence curve.

\begin{equation}
    R_N \sim -(1-\Tilde{T})^{-\alpha}
\end{equation}
or, equivalently,
\begin{equation}
    \ln{|R_N|} = -\alpha \ln{|1-\Tilde{T}|} + \beta
\end{equation}
near the critical point. Fig. \ref{CurCoexistenceSWSHLog} shows the discrete points and the fitted curves. Along the coexistence swollen phase curve, we get the following fitting result,
\begin{equation}\label{CSW}
    \ln{|R_N|} = -2.00902 \ln{|1-\Tilde{T}|} -5.01464
\end{equation}
The result along the coexistence shrunken phase curve is,
\begin{equation}\label{CSH}
    \ln{|R_N|} = -1.99123 \ln{|1-\Tilde{T}|} -4.7542
\end{equation}
Comparing Eqs. \eqref{CSW} and \eqref{CSH}, we can deduce that the slope of $\ln{|R_N|}$ vs. $\ln{|1-\Tilde{T}|}$ is -2. Thus, the critical exponent associated with scalar curvature is 2. Also, $R_N$ as a function of $\Tilde{T}$ is given by,
 
\begin{equation}\label{RelationCur}
    R_N \sim -(1-\Tilde{T})^{-2}
\end{equation}
The critical exponent associated with curvature from the relation \eqref{RelationCur} is 2. Assuming that the critical exponent associated with the correlation length $\xi$ is $\nu$ i.e $\xi \sim (1-\Tilde{T})^{-\nu}$, we may obtain the following relationship between curvature and correlation length,

\begin{equation}\label{FinalReln}
    R_N \sim -\xi^{2/\nu}
\end{equation}
The fact that the critical exponent associated with the Riemannian curvature is 2 and the scaling of the curvature with the correlation length is given by Eq. \eqref{FinalReln} establish that phase transition in gels indeed belongs to the same universality class as van der Waals fluids and black hole systems in line with \cite{wei2019repulsive,wei2019ruppeiner,sarkar2014}. 

\section{Conclusion}\label{Conclusion}

In view of various practical applications, discontinuous volumetric phase transition in gels is of interest to both scientific and industrial communities. A substantive characterization of phase transition based on an understanding of the microstructure is thus crucial. In the present article, we have studied phase separation in the light of a Landau-Ginzburg theory.
We have proposed a modification to the free energy to eliminate unphysical non-convexity. Such a modification removes the necessity of the Maxwell construction in first order phase transitions.
Further, we have used Ruppeiner's thermodynamic approach to probe the microstructure of a polyacrylamide gel during the transition regime. The divergence points in the Ricci scalar curvature indicate diverging correlation length and hence large density fluctuations. This implies phase segregation and a heterogeneous microstructure during transition. Further, the uniform negative sign of the curvature indicates no drastic change in the molecular interactions for the range of temperatures considered. Finally, we have numerically calculated the critical exponent associated with the scalar curvature and established a scaling law between curvature and correlation length. We have thus theoretically confirmed that this phase transition belongs to the same universality class as gas-liquid phase transition in van der Waals fluids. We may also draw inferences on the gel microstructure across transitions and across critical points. While the microstructure is phase-segregated during transition and at critical point, no such conclusion can be reached in the  supercritical regime with no singularities in curvature post the critical point. In the supercritical regime, from the present approach, we may merely assume that the nature of interactions does not change and that the microstructure is homogeneous, pending a more detailed study where curvature needs to be analyzed at even higher temperatures with more accurate notions of specific heat in line with \cite{bolmatov2013thermodynamic}.
The modification to the free energy proposed in the present article renders the energy convex, but not strongly convex. A more general approach might be possible via large deviation functions \cite{TOUCHETTE20091} which may remove the unphysical aspects of metric being zero and subsequently, the curvature being infinity. In our future work, we may explore this aspect further. Also we would exploit our observations to design efficient supercritical systems for biomedical and structural applications.

\providecommand{\noopsort}[1]{}\providecommand{\singleletter}[1]{#1}%


\begin{thebibliography}{61}%
\makeatletter
\providecommand \@ifxundefined [1]{%
 \@ifx{#1\undefined}
}%
\providecommand \@ifnum [1]{%
 \ifnum #1\expandafter \@firstoftwo
 \else \expandafter \@secondoftwo
 \fi
}%
\providecommand \@ifx [1]{%
 \ifx #1\expandafter \@firstoftwo
 \else \expandafter \@secondoftwo
 \fi
}%
\providecommand \natexlab [1]{#1}%
\providecommand \enquote  [1]{``#1''}%
\providecommand \bibnamefont  [1]{#1}%
\providecommand \bibfnamefont [1]{#1}%
\providecommand \citenamefont [1]{#1}%
\providecommand \href@noop [0]{\@secondoftwo}%
\providecommand \href [0]{\begingroup \@sanitize@url \@href}%
\providecommand \@href[1]{\@@startlink{#1}\@@href}%
\providecommand \@@href[1]{\endgroup#1\@@endlink}%
\providecommand \@sanitize@url [0]{\catcode `\\12\catcode `\$12\catcode `\&12\catcode `\#12\catcode `\^12\catcode `\_12\catcode `\%12\relax}%
\providecommand \@@startlink[1]{}%
\providecommand \@@endlink[0]{}%
\providecommand \url  [0]{\begingroup\@sanitize@url \@url }%
\providecommand \@url [1]{\endgroup\@href {#1}{\urlprefix }}%
\providecommand \urlprefix  [0]{URL }%
\providecommand \Eprint [0]{\href }%
\providecommand \doibase [0]{https://doi.org/}%
\providecommand \selectlanguage [0]{\@gobble}%
\providecommand \bibinfo  [0]{\@secondoftwo}%
\providecommand \bibfield  [0]{\@secondoftwo}%
\providecommand \translation [1]{[#1]}%
\providecommand \BibitemOpen [0]{}%
\providecommand \bibitemStop [0]{}%
\providecommand \bibitemNoStop [0]{.\EOS\space}%
\providecommand \EOS [0]{\spacefactor3000\relax}%
\providecommand \BibitemShut  [1]{\csname bibitem#1\endcsname}%
\let\auto@bib@innerbib\@empty
\bibitem [{\citenamefont {Koetting}\ \emph {et~al.}(2015)\citenamefont {Koetting}, \citenamefont {Peters}, \citenamefont {Steichen},\ and\ \citenamefont {Peppas}}]{koetting2015stimulus}%
  \BibitemOpen
  \bibfield  {author} {\bibinfo {author} {\bibfnamefont {M.~C.}\ \bibnamefont {Koetting}}, \bibinfo {author} {\bibfnamefont {J.~T.}\ \bibnamefont {Peters}}, \bibinfo {author} {\bibfnamefont {S.~D.}\ \bibnamefont {Steichen}},\ and\ \bibinfo {author} {\bibfnamefont {N.~A.}\ \bibnamefont {Peppas}},\ }\bibfield  {title} {\bibinfo {title} {Stimulus-responsive hydrogels: Theory, modern advances, and applications},\ }\href@noop {} {\bibfield  {journal} {\bibinfo  {journal} {Materials Science and Engineering: R: Reports}\ }\textbf {\bibinfo {volume} {93}},\ \bibinfo {pages} {1} (\bibinfo {year} {2015})}\BibitemShut {NoStop}%
\bibitem [{\citenamefont {Kamata}\ \emph {et~al.}(2015)\citenamefont {Kamata}, \citenamefont {Li}, \citenamefont {Chung},\ and\ \citenamefont {Sakai}}]{kamata2015design}%
  \BibitemOpen
  \bibfield  {author} {\bibinfo {author} {\bibfnamefont {H.}~\bibnamefont {Kamata}}, \bibinfo {author} {\bibfnamefont {X.}~\bibnamefont {Li}}, \bibinfo {author} {\bibfnamefont {U.-i.}\ \bibnamefont {Chung}},\ and\ \bibinfo {author} {\bibfnamefont {T.}~\bibnamefont {Sakai}},\ }\bibfield  {title} {\bibinfo {title} {Design of hydrogels for biomedical applications},\ }\href@noop {} {\bibfield  {journal} {\bibinfo  {journal} {Advanced healthcare materials}\ }\textbf {\bibinfo {volume} {4}},\ \bibinfo {pages} {2360} (\bibinfo {year} {2015})}\BibitemShut {NoStop}%
\bibitem [{\citenamefont {Shibayama}\ and\ \citenamefont {Tanaka}(1993)}]{shibayama1993volume}%
  \BibitemOpen
  \bibfield  {author} {\bibinfo {author} {\bibfnamefont {M.}~\bibnamefont {Shibayama}}\ and\ \bibinfo {author} {\bibfnamefont {T.}~\bibnamefont {Tanaka}},\ }\bibfield  {title} {\bibinfo {title} {Volume phase transition and related phenomena of polymer gels},\ }\href@noop {} {\bibfield  {journal} {\bibinfo  {journal} {Responsive gels: volume transitions I}\ ,\ \bibinfo {pages} {1}} (\bibinfo {year} {1993})}\BibitemShut {NoStop}%
\bibitem [{\citenamefont {Tanaka}\ \emph {et~al.}(1977)\citenamefont {Tanaka}, \citenamefont {Ishiwata},\ and\ \citenamefont {Ishimoto}}]{tanaka1977critical}%
  \BibitemOpen
  \bibfield  {author} {\bibinfo {author} {\bibfnamefont {T.}~\bibnamefont {Tanaka}}, \bibinfo {author} {\bibfnamefont {S.}~\bibnamefont {Ishiwata}},\ and\ \bibinfo {author} {\bibfnamefont {C.}~\bibnamefont {Ishimoto}},\ }\bibfield  {title} {\bibinfo {title} {Critical behavior of density fluctuations in gels},\ }\href@noop {} {\bibfield  {journal} {\bibinfo  {journal} {Physical Review Letters}\ }\textbf {\bibinfo {volume} {38}},\ \bibinfo {pages} {771} (\bibinfo {year} {1977})}\BibitemShut {NoStop}%
\bibitem [{\citenamefont {Tanaka}(1978)}]{tanaka1978collapse}%
  \BibitemOpen
  \bibfield  {author} {\bibinfo {author} {\bibfnamefont {T.}~\bibnamefont {Tanaka}},\ }\bibfield  {title} {\bibinfo {title} {Collapse of gels and the critical endpoint},\ }\href@noop {} {\bibfield  {journal} {\bibinfo  {journal} {Physical review letters}\ }\textbf {\bibinfo {volume} {40}},\ \bibinfo {pages} {820} (\bibinfo {year} {1978})}\BibitemShut {NoStop}%
\bibitem [{\citenamefont {Du{\v{s}}ek}\ and\ \citenamefont {Patterson}(1968)}]{duvsek1968transition}%
  \BibitemOpen
  \bibfield  {author} {\bibinfo {author} {\bibfnamefont {K.}~\bibnamefont {Du{\v{s}}ek}}\ and\ \bibinfo {author} {\bibfnamefont {D.}~\bibnamefont {Patterson}},\ }\bibfield  {title} {\bibinfo {title} {Transition in swollen polymer networks induced by intramolecular condensation},\ }\href@noop {} {\bibfield  {journal} {\bibinfo  {journal} {Journal of Polymer Science Part A-2: Polymer Physics}\ }\textbf {\bibinfo {volume} {6}},\ \bibinfo {pages} {1209} (\bibinfo {year} {1968})}\BibitemShut {NoStop}%
\bibitem [{\citenamefont {Amiya}\ and\ \citenamefont {Tanaka}(1987)}]{amiya1987phase}%
  \BibitemOpen
  \bibfield  {author} {\bibinfo {author} {\bibfnamefont {T.}~\bibnamefont {Amiya}}\ and\ \bibinfo {author} {\bibfnamefont {T.}~\bibnamefont {Tanaka}},\ }\bibfield  {title} {\bibinfo {title} {Phase transitions in crosslinked gels of natural polymers},\ }\href@noop {} {\bibfield  {journal} {\bibinfo  {journal} {Macromolecules}\ }\textbf {\bibinfo {volume} {20}},\ \bibinfo {pages} {1162} (\bibinfo {year} {1987})}\BibitemShut {NoStop}%
\bibitem [{\citenamefont {Katayama}\ and\ \citenamefont {Ohata}(1985)}]{katayama1985phase}%
  \BibitemOpen
  \bibfield  {author} {\bibinfo {author} {\bibfnamefont {S.}~\bibnamefont {Katayama}}\ and\ \bibinfo {author} {\bibfnamefont {A.}~\bibnamefont {Ohata}},\ }\bibfield  {title} {\bibinfo {title} {Phase transition of a cationic gel},\ }\href@noop {} {\bibfield  {journal} {\bibinfo  {journal} {Macromolecules}\ }\textbf {\bibinfo {volume} {18}},\ \bibinfo {pages} {2781} (\bibinfo {year} {1985})}\BibitemShut {NoStop}%
\bibitem [{\citenamefont {Hirokawa}\ and\ \citenamefont {Tanaka}(1984)}]{hirokawa1984volume}%
  \BibitemOpen
  \bibfield  {author} {\bibinfo {author} {\bibfnamefont {Y.}~\bibnamefont {Hirokawa}}\ and\ \bibinfo {author} {\bibfnamefont {T.}~\bibnamefont {Tanaka}},\ }\bibfield  {title} {\bibinfo {title} {Volume phase transition in a non-ionic gel},\ }in\ \href@noop {} {\emph {\bibinfo {booktitle} {AIP Conference Proceedings}}},\ Vol.\ \bibinfo {volume} {107}\ (\bibinfo {organization} {American Institute of Physics},\ \bibinfo {year} {1984})\ pp.\ \bibinfo {pages} {203--208}\BibitemShut {NoStop}%
\bibitem [{\citenamefont {Hirotsu}\ \emph {et~al.}(1987)\citenamefont {Hirotsu}, \citenamefont {Hirokawa},\ and\ \citenamefont {Tanaka}}]{hirotsu1987volume}%
  \BibitemOpen
  \bibfield  {author} {\bibinfo {author} {\bibfnamefont {S.}~\bibnamefont {Hirotsu}}, \bibinfo {author} {\bibfnamefont {Y.}~\bibnamefont {Hirokawa}},\ and\ \bibinfo {author} {\bibfnamefont {T.}~\bibnamefont {Tanaka}},\ }\bibfield  {title} {\bibinfo {title} {Volume-phase transitions of ionized n-isopropylacrylamide gels},\ }\href@noop {} {\bibfield  {journal} {\bibinfo  {journal} {The Journal of chemical physics}\ }\textbf {\bibinfo {volume} {87}},\ \bibinfo {pages} {1392} (\bibinfo {year} {1987})}\BibitemShut {NoStop}%
\bibitem [{\citenamefont {Otake}\ \emph {et~al.}(1988)\citenamefont {Otake}, \citenamefont {Tsuji}, \citenamefont {Konno},\ and\ \citenamefont {Saito}}]{otake1988preparation}%
  \BibitemOpen
  \bibfield  {author} {\bibinfo {author} {\bibfnamefont {K.}~\bibnamefont {Otake}}, \bibinfo {author} {\bibfnamefont {T.}~\bibnamefont {Tsuji}}, \bibinfo {author} {\bibfnamefont {M.}~\bibnamefont {Konno}},\ and\ \bibinfo {author} {\bibfnamefont {S.}~\bibnamefont {Saito}},\ }\bibfield  {title} {\bibinfo {title} {Preparation of a new hydrogel and porous glass composite membrane},\ }\href@noop {} {\bibfield  {journal} {\bibinfo  {journal} {Journal of chemical engineering of Japan}\ }\textbf {\bibinfo {volume} {21}},\ \bibinfo {pages} {443} (\bibinfo {year} {1988})}\BibitemShut {NoStop}%
\bibitem [{\citenamefont {Inomata}\ \emph {et~al.}(1990)\citenamefont {Inomata}, \citenamefont {Goto},\ and\ \citenamefont {Saito}}]{inomata1990phase}%
  \BibitemOpen
  \bibfield  {author} {\bibinfo {author} {\bibfnamefont {H.}~\bibnamefont {Inomata}}, \bibinfo {author} {\bibfnamefont {S.}~\bibnamefont {Goto}},\ and\ \bibinfo {author} {\bibfnamefont {S.}~\bibnamefont {Saito}},\ }\bibfield  {title} {\bibinfo {title} {Phase transition of n-substituted acrylamide gels},\ }\href@noop {} {\bibfield  {journal} {\bibinfo  {journal} {Macromolecules}\ }\textbf {\bibinfo {volume} {23}},\ \bibinfo {pages} {4887} (\bibinfo {year} {1990})}\BibitemShut {NoStop}%
\bibitem [{\citenamefont {Otake}\ \emph {et~al.}(1990)\citenamefont {Otake}, \citenamefont {Inomata}, \citenamefont {Konno},\ and\ \citenamefont {Saito}}]{otake1990thermal}%
  \BibitemOpen
  \bibfield  {author} {\bibinfo {author} {\bibfnamefont {K.}~\bibnamefont {Otake}}, \bibinfo {author} {\bibfnamefont {H.}~\bibnamefont {Inomata}}, \bibinfo {author} {\bibfnamefont {M.}~\bibnamefont {Konno}},\ and\ \bibinfo {author} {\bibfnamefont {S.}~\bibnamefont {Saito}},\ }\bibfield  {title} {\bibinfo {title} {Thermal analysis of the volume phase transition with n-isopropylacrylamide gels},\ }\href@noop {} {\bibfield  {journal} {\bibinfo  {journal} {Macromolecules}\ }\textbf {\bibinfo {volume} {23}},\ \bibinfo {pages} {283} (\bibinfo {year} {1990})}\BibitemShut {NoStop}%
\bibitem [{\citenamefont {Siegel}\ \emph {et~al.}(1988)\citenamefont {Siegel}, \citenamefont {Falamarzian}, \citenamefont {Firestone},\ and\ \citenamefont {Moxley}}]{siegel1988ph}%
  \BibitemOpen
  \bibfield  {author} {\bibinfo {author} {\bibfnamefont {R.~A.}\ \bibnamefont {Siegel}}, \bibinfo {author} {\bibfnamefont {M.}~\bibnamefont {Falamarzian}}, \bibinfo {author} {\bibfnamefont {B.~A.}\ \bibnamefont {Firestone}},\ and\ \bibinfo {author} {\bibfnamefont {B.~C.}\ \bibnamefont {Moxley}},\ }\bibfield  {title} {\bibinfo {title} {ph-controlled release from hydrophobic/polyelectrolyte copolymer hydrogels},\ }\href@noop {} {\bibfield  {journal} {\bibinfo  {journal} {Journal of Controlled Release}\ }\textbf {\bibinfo {volume} {8}},\ \bibinfo {pages} {179} (\bibinfo {year} {1988})}\BibitemShut {NoStop}%
\bibitem [{\citenamefont {Hirokawa}\ \emph {et~al.}(1984)\citenamefont {Hirokawa}, \citenamefont {Tanaka},\ and\ \citenamefont {Katayama}}]{hirokawa1984microbial}%
  \BibitemOpen
  \bibfield  {author} {\bibinfo {author} {\bibfnamefont {Y.}~\bibnamefont {Hirokawa}}, \bibinfo {author} {\bibfnamefont {T.}~\bibnamefont {Tanaka}},\ and\ \bibinfo {author} {\bibfnamefont {S.}~\bibnamefont {Katayama}},\ }\href@noop {} {\bibinfo {title} {Microbial adhesion and aggregation" ed. kc marshall}} (\bibinfo {year} {1984})\BibitemShut {NoStop}%
\bibitem [{\citenamefont {Ricka}\ and\ \citenamefont {Tanaka}(1984)}]{ricka1984swelling}%
  \BibitemOpen
  \bibfield  {author} {\bibinfo {author} {\bibfnamefont {J.}~\bibnamefont {Ricka}}\ and\ \bibinfo {author} {\bibfnamefont {T.}~\bibnamefont {Tanaka}},\ }\bibfield  {title} {\bibinfo {title} {Swelling of ionic gels: quantitative performance of the donnan theory},\ }\href@noop {} {\bibfield  {journal} {\bibinfo  {journal} {Macromolecules}\ }\textbf {\bibinfo {volume} {17}},\ \bibinfo {pages} {2916} (\bibinfo {year} {1984})}\BibitemShut {NoStop}%
\bibitem [{\citenamefont {Ohmine}\ and\ \citenamefont {Tanaka}(1982)}]{ohmine1982salt}%
  \BibitemOpen
  \bibfield  {author} {\bibinfo {author} {\bibfnamefont {I.}~\bibnamefont {Ohmine}}\ and\ \bibinfo {author} {\bibfnamefont {T.}~\bibnamefont {Tanaka}},\ }\bibfield  {title} {\bibinfo {title} {Salt effects on the phase transition of ionic gels},\ }\href@noop {} {\bibfield  {journal} {\bibinfo  {journal} {The Journal of Chemical Physics}\ }\textbf {\bibinfo {volume} {77}},\ \bibinfo {pages} {5725} (\bibinfo {year} {1982})}\BibitemShut {NoStop}%
\bibitem [{\citenamefont {Siegel}\ and\ \citenamefont {Firestone}(1988)}]{siegel1988ph2}%
  \BibitemOpen
  \bibfield  {author} {\bibinfo {author} {\bibfnamefont {R.~A.}\ \bibnamefont {Siegel}}\ and\ \bibinfo {author} {\bibfnamefont {B.~A.}\ \bibnamefont {Firestone}},\ }\bibfield  {title} {\bibinfo {title} {ph-dependent equilibrium swelling properties of hydrophobic polyelectrolyte copolymer gels},\ }\href@noop {} {\bibfield  {journal} {\bibinfo  {journal} {Macromolecules}\ }\textbf {\bibinfo {volume} {21}},\ \bibinfo {pages} {3254} (\bibinfo {year} {1988})}\BibitemShut {NoStop}%
\bibitem [{\citenamefont {Osada}(2005)}]{osada2005conversion}%
  \BibitemOpen
  \bibfield  {author} {\bibinfo {author} {\bibfnamefont {Y.}~\bibnamefont {Osada}},\ }\bibfield  {title} {\bibinfo {title} {Conversion of chemical into mechanical energy by synthetic polymers (chemomechanical systems)},\ }in\ \href@noop {} {\emph {\bibinfo {booktitle} {Polymer Physics}}}\ (\bibinfo  {publisher} {Springer},\ \bibinfo {year} {2005})\ pp.\ \bibinfo {pages} {1--46}\BibitemShut {NoStop}%
\bibitem [{\citenamefont {Tanaka}\ \emph {et~al.}(1982)\citenamefont {Tanaka}, \citenamefont {Nishio}, \citenamefont {Sun},\ and\ \citenamefont {Ueno-Nishio}}]{tanaka1982collapse}%
  \BibitemOpen
  \bibfield  {author} {\bibinfo {author} {\bibfnamefont {T.}~\bibnamefont {Tanaka}}, \bibinfo {author} {\bibfnamefont {I.}~\bibnamefont {Nishio}}, \bibinfo {author} {\bibfnamefont {S.-T.}\ \bibnamefont {Sun}},\ and\ \bibinfo {author} {\bibfnamefont {S.}~\bibnamefont {Ueno-Nishio}},\ }\bibfield  {title} {\bibinfo {title} {Collapse of gels in an electric field},\ }\href@noop {} {\bibfield  {journal} {\bibinfo  {journal} {Science}\ }\textbf {\bibinfo {volume} {218}},\ \bibinfo {pages} {467} (\bibinfo {year} {1982})}\BibitemShut {NoStop}%
\bibitem [{\citenamefont {Giannetti}\ \emph {et~al.}(1988)\citenamefont {Giannetti}, \citenamefont {Hirose}, \citenamefont {Hirokawa},\ and\ \citenamefont {Tanaka}}]{giannetti1988molecular}%
  \BibitemOpen
  \bibfield  {author} {\bibinfo {author} {\bibfnamefont {G.}~\bibnamefont {Giannetti}}, \bibinfo {author} {\bibfnamefont {Y.}~\bibnamefont {Hirose}}, \bibinfo {author} {\bibfnamefont {Y.}~\bibnamefont {Hirokawa}},\ and\ \bibinfo {author} {\bibfnamefont {T.}~\bibnamefont {Tanaka}},\ }\href@noop {} {\bibinfo {title} {Molecular electronic devices}} (\bibinfo {year} {1988})\BibitemShut {NoStop}%
\bibitem [{\citenamefont {Flory}(1953)}]{flory1953principles}%
  \BibitemOpen
  \bibfield  {author} {\bibinfo {author} {\bibfnamefont {P.~J.}\ \bibnamefont {Flory}},\ }\href@noop {} {\emph {\bibinfo {title} {Principles of polymer chemistry}}}\ (\bibinfo  {publisher} {Cornell university press},\ \bibinfo {year} {1953})\BibitemShut {NoStop}%
\bibitem [{\citenamefont {Cahn}\ and\ \citenamefont {Hilliard}(1958)}]{cahn1958free}%
  \BibitemOpen
  \bibfield  {author} {\bibinfo {author} {\bibfnamefont {J.~W.}\ \bibnamefont {Cahn}}\ and\ \bibinfo {author} {\bibfnamefont {J.~E.}\ \bibnamefont {Hilliard}},\ }\bibfield  {title} {\bibinfo {title} {Free energy of a nonuniform system. i. interfacial free energy},\ }\href@noop {} {\bibfield  {journal} {\bibinfo  {journal} {The Journal of chemical physics}\ }\textbf {\bibinfo {volume} {28}},\ \bibinfo {pages} {258} (\bibinfo {year} {1958})}\BibitemShut {NoStop}%
\bibitem [{\citenamefont {Cahn}\ and\ \citenamefont {Hilliard}(1959)}]{cahn1959free}%
  \BibitemOpen
  \bibfield  {author} {\bibinfo {author} {\bibfnamefont {J.~W.}\ \bibnamefont {Cahn}}\ and\ \bibinfo {author} {\bibfnamefont {J.~E.}\ \bibnamefont {Hilliard}},\ }\bibfield  {title} {\bibinfo {title} {Free energy of a nonuniform system. iii. nucleation in a two-component incompressible fluid},\ }\href@noop {} {\bibfield  {journal} {\bibinfo  {journal} {The Journal of chemical physics}\ }\textbf {\bibinfo {volume} {31}},\ \bibinfo {pages} {688} (\bibinfo {year} {1959})}\BibitemShut {NoStop}%
\bibitem [{\citenamefont {Nauman}\ and\ \citenamefont {Balsara}(1989)}]{nauman1989phase}%
  \BibitemOpen
  \bibfield  {author} {\bibinfo {author} {\bibfnamefont {E.}~\bibnamefont {Nauman}}\ and\ \bibinfo {author} {\bibfnamefont {N.~P.}\ \bibnamefont {Balsara}},\ }\bibfield  {title} {\bibinfo {title} {Phase equilibria and the landau—ginzburg functional},\ }\href@noop {} {\bibfield  {journal} {\bibinfo  {journal} {Fluid Phase Equilibria}\ }\textbf {\bibinfo {volume} {45}},\ \bibinfo {pages} {229} (\bibinfo {year} {1989})}\BibitemShut {NoStop}%
\bibitem [{\citenamefont {Binder}(1987{\natexlab{a}})}]{binder1987theory}%
  \BibitemOpen
  \bibfield  {author} {\bibinfo {author} {\bibfnamefont {K.}~\bibnamefont {Binder}},\ }\bibfield  {title} {\bibinfo {title} {Theory of first-order phase transitions},\ }\href@noop {} {\bibfield  {journal} {\bibinfo  {journal} {Reports on progress in physics}\ }\textbf {\bibinfo {volume} {50}},\ \bibinfo {pages} {783} (\bibinfo {year} {1987}{\natexlab{a}})}\BibitemShut {NoStop}%
\bibitem [{\citenamefont {Binder}(1987{\natexlab{b}})}]{binder1987dynamics}%
  \BibitemOpen
  \bibfield  {author} {\bibinfo {author} {\bibfnamefont {K.}~\bibnamefont {Binder}},\ }\bibfield  {title} {\bibinfo {title} {Dynamics of phase separation and critical phenomena in polymer mixtures},\ }\href@noop {} {\bibfield  {journal} {\bibinfo  {journal} {Colloid and Polymer Science}\ }\textbf {\bibinfo {volume} {265}},\ \bibinfo {pages} {273} (\bibinfo {year} {1987}{\natexlab{b}})}\BibitemShut {NoStop}%
\bibitem [{\citenamefont {Ariyapadi}\ and\ \citenamefont {Nauman}(1990)}]{ariyapadi1990gradient}%
  \BibitemOpen
  \bibfield  {author} {\bibinfo {author} {\bibfnamefont {M.}~\bibnamefont {Ariyapadi}}\ and\ \bibinfo {author} {\bibfnamefont {E.}~\bibnamefont {Nauman}},\ }\bibfield  {title} {\bibinfo {title} {Gradient energy parameters for polymer--polymer--solvent systems and their application to spinodal decomposition in true ternary systems},\ }\href@noop {} {\bibfield  {journal} {\bibinfo  {journal} {Journal of Polymer Science Part B: Polymer Physics}\ }\textbf {\bibinfo {volume} {28}},\ \bibinfo {pages} {2395} (\bibinfo {year} {1990})}\BibitemShut {NoStop}%
\bibitem [{\citenamefont {He}\ \emph {et~al.}(1996)\citenamefont {He}, \citenamefont {Kwak},\ and\ \citenamefont {Nauman}}]{he1996phase}%
  \BibitemOpen
  \bibfield  {author} {\bibinfo {author} {\bibfnamefont {D.~Q.}\ \bibnamefont {He}}, \bibinfo {author} {\bibfnamefont {S.}~\bibnamefont {Kwak}},\ and\ \bibinfo {author} {\bibfnamefont {E.~B.}\ \bibnamefont {Nauman}},\ }\bibfield  {title} {\bibinfo {title} {On phase equilibria, interfacial tension and phase growth in ternary polymer blends},\ }\href@noop {} {\bibfield  {journal} {\bibinfo  {journal} {Macromolecular theory and simulations}\ }\textbf {\bibinfo {volume} {5}},\ \bibinfo {pages} {801} (\bibinfo {year} {1996})}\BibitemShut {NoStop}%
\bibitem [{\citenamefont {Nauman}\ and\ \citenamefont {He}(2001)}]{NAUMAN2001}%
  \BibitemOpen
  \bibfield  {author} {\bibinfo {author} {\bibfnamefont {E.}~\bibnamefont {Nauman}}\ and\ \bibinfo {author} {\bibfnamefont {D.~Q.}\ \bibnamefont {He}},\ }\bibfield  {title} {\bibinfo {title} {Nonlinear diffusion and phase separation},\ }\href {https://doi.org/https://doi.org/10.1016/S0009-2509(01)00005-7} {\bibfield  {journal} {\bibinfo  {journal} {Chemical Engineering Science}\ }\textbf {\bibinfo {volume} {56}},\ \bibinfo {pages} {1999} (\bibinfo {year} {2001})}\BibitemShut {NoStop}%
\bibitem [{\citenamefont {Li}\ and\ \citenamefont {Tanaka}(1992)}]{li1992phase}%
  \BibitemOpen
  \bibfield  {author} {\bibinfo {author} {\bibfnamefont {Y.}~\bibnamefont {Li}}\ and\ \bibinfo {author} {\bibfnamefont {T.}~\bibnamefont {Tanaka}},\ }\bibfield  {title} {\bibinfo {title} {Phase transitions of gels},\ }\href@noop {} {\bibfield  {journal} {\bibinfo  {journal} {Annual Review of Materials Science}\ }\textbf {\bibinfo {volume} {22}},\ \bibinfo {pages} {243} (\bibinfo {year} {1992})}\BibitemShut {NoStop}%
\bibitem [{\citenamefont {Hochberg}\ \emph {et~al.}(1979)\citenamefont {Hochberg}, \citenamefont {Tanaka},\ and\ \citenamefont {Nicoli}}]{hochberg1979spinodal}%
  \BibitemOpen
  \bibfield  {author} {\bibinfo {author} {\bibfnamefont {A.}~\bibnamefont {Hochberg}}, \bibinfo {author} {\bibfnamefont {T.}~\bibnamefont {Tanaka}},\ and\ \bibinfo {author} {\bibfnamefont {D.}~\bibnamefont {Nicoli}},\ }\bibfield  {title} {\bibinfo {title} {Spinodal line and critical point of an acrylamide gel},\ }\href@noop {} {\bibfield  {journal} {\bibinfo  {journal} {Physical Review Letters}\ }\textbf {\bibinfo {volume} {43}},\ \bibinfo {pages} {217} (\bibinfo {year} {1979})}\BibitemShut {NoStop}%
\bibitem [{\citenamefont {Hirotsu}(1987)}]{hirotsu1987phase}%
  \BibitemOpen
  \bibfield  {author} {\bibinfo {author} {\bibfnamefont {S.}~\bibnamefont {Hirotsu}},\ }\bibfield  {title} {\bibinfo {title} {Phase transition of a polymer gel in pure and mixed solvent media},\ }\href@noop {} {\bibfield  {journal} {\bibinfo  {journal} {Journal of the Physical Society of Japan}\ }\textbf {\bibinfo {volume} {56}},\ \bibinfo {pages} {233} (\bibinfo {year} {1987})}\BibitemShut {NoStop}%
\bibitem [{\citenamefont {Tokita}\ and\ \citenamefont {Tanaka}(1991)}]{tokita1991friction}%
  \BibitemOpen
  \bibfield  {author} {\bibinfo {author} {\bibfnamefont {M.}~\bibnamefont {Tokita}}\ and\ \bibinfo {author} {\bibfnamefont {T.}~\bibnamefont {Tanaka}},\ }\bibfield  {title} {\bibinfo {title} {Friction coefficient of polymer networks of gels},\ }\href@noop {} {\bibfield  {journal} {\bibinfo  {journal} {The Journal of chemical physics}\ }\textbf {\bibinfo {volume} {95}},\ \bibinfo {pages} {4613} (\bibinfo {year} {1991})}\BibitemShut {NoStop}%
\bibitem [{\citenamefont {Li}\ and\ \citenamefont {Tanaka}(1989)}]{li1989study}%
  \BibitemOpen
  \bibfield  {author} {\bibinfo {author} {\bibfnamefont {Y.}~\bibnamefont {Li}}\ and\ \bibinfo {author} {\bibfnamefont {T.}~\bibnamefont {Tanaka}},\ }\bibfield  {title} {\bibinfo {title} {Study of the universality class of the gel network system},\ }\href@noop {} {\bibfield  {journal} {\bibinfo  {journal} {The Journal of chemical physics}\ }\textbf {\bibinfo {volume} {90}},\ \bibinfo {pages} {5161} (\bibinfo {year} {1989})}\BibitemShut {NoStop}%
\bibitem [{\citenamefont {Ruppeiner}(1995)}]{ruppeiner1995riemannian}%
  \BibitemOpen
  \bibfield  {author} {\bibinfo {author} {\bibfnamefont {G.}~\bibnamefont {Ruppeiner}},\ }\bibfield  {title} {\bibinfo {title} {Riemannian geometry in thermodynamic fluctuation theory},\ }\href@noop {} {\bibfield  {journal} {\bibinfo  {journal} {Reviews of Modern Physics}\ }\textbf {\bibinfo {volume} {67}},\ \bibinfo {pages} {605} (\bibinfo {year} {1995})}\BibitemShut {NoStop}%
\bibitem [{\citenamefont {Das}\ \emph {et~al.}(2022)\citenamefont {Das}, \citenamefont {Raza},\ and\ \citenamefont {Roy}}]{das2022geometric}%
  \BibitemOpen
  \bibfield  {author} {\bibinfo {author} {\bibfnamefont {S.}~\bibnamefont {Das}}, \bibinfo {author} {\bibfnamefont {A.}~\bibnamefont {Raza}},\ and\ \bibinfo {author} {\bibfnamefont {D.}~\bibnamefont {Roy}},\ }\bibfield  {title} {\bibinfo {title} {Geometric thermodynamics of strain-induced crystallization in polymers},\ }\href {https://doi.org/10.1103/PhysRevE.106.015005} {\bibfield  {journal} {\bibinfo  {journal} {Phys. Rev. E}\ }\textbf {\bibinfo {volume} {106}},\ \bibinfo {pages} {015005} (\bibinfo {year} {2022})}\BibitemShut {NoStop}%
\bibitem [{\citenamefont {Ruppeiner}(2010)}]{ruppeiner2010thermodynamic}%
  \BibitemOpen
  \bibfield  {author} {\bibinfo {author} {\bibfnamefont {G.}~\bibnamefont {Ruppeiner}},\ }\bibfield  {title} {\bibinfo {title} {Thermodynamic curvature measures interactions},\ }\href@noop {} {\bibfield  {journal} {\bibinfo  {journal} {American Journal of Physics}\ }\textbf {\bibinfo {volume} {78}},\ \bibinfo {pages} {1170} (\bibinfo {year} {2010})}\BibitemShut {NoStop}%
\bibitem [{\citenamefont {Ruppeiner}\ and\ \citenamefont {Bellucci}(2015)}]{ruppeiner2015thermodynamic}%
  \BibitemOpen
  \bibfield  {author} {\bibinfo {author} {\bibfnamefont {G.}~\bibnamefont {Ruppeiner}}\ and\ \bibinfo {author} {\bibfnamefont {S.}~\bibnamefont {Bellucci}},\ }\bibfield  {title} {\bibinfo {title} {Thermodynamic curvature for a two-parameter spin model with frustration},\ }\href@noop {} {\bibfield  {journal} {\bibinfo  {journal} {Physical Review E}\ }\textbf {\bibinfo {volume} {91}},\ \bibinfo {pages} {012116} (\bibinfo {year} {2015})}\BibitemShut {NoStop}%
\bibitem [{\citenamefont {Ruppeiner}(2008)}]{ruppeiner2008thermodynamic}%
  \BibitemOpen
  \bibfield  {author} {\bibinfo {author} {\bibfnamefont {G.}~\bibnamefont {Ruppeiner}},\ }\bibfield  {title} {\bibinfo {title} {Thermodynamic curvature and phase transitions in kerr-newman black holes},\ }\href@noop {} {\bibfield  {journal} {\bibinfo  {journal} {Physical Review D}\ }\textbf {\bibinfo {volume} {78}},\ \bibinfo {pages} {024016} (\bibinfo {year} {2008})}\BibitemShut {NoStop}%
\bibitem [{\citenamefont {Ruppeiner}\ and\ \citenamefont {Seftas}(2020)}]{ruppeiner2020thermodynamic}%
  \BibitemOpen
  \bibfield  {author} {\bibinfo {author} {\bibfnamefont {G.}~\bibnamefont {Ruppeiner}}\ and\ \bibinfo {author} {\bibfnamefont {A.}~\bibnamefont {Seftas}},\ }\bibfield  {title} {\bibinfo {title} {Thermodynamic curvature of the binary van der waals fluid},\ }\href@noop {} {\bibfield  {journal} {\bibinfo  {journal} {Entropy}\ }\textbf {\bibinfo {volume} {22}},\ \bibinfo {pages} {1208} (\bibinfo {year} {2020})}\BibitemShut {NoStop}%
\bibitem [{\citenamefont {Ruppeiner}\ \emph {et~al.}(2021)\citenamefont {Ruppeiner}, \citenamefont {Mausbach},\ and\ \citenamefont {May}}]{ruppeiner2021thermodynamic}%
  \BibitemOpen
  \bibfield  {author} {\bibinfo {author} {\bibfnamefont {G.}~\bibnamefont {Ruppeiner}}, \bibinfo {author} {\bibfnamefont {P.}~\bibnamefont {Mausbach}},\ and\ \bibinfo {author} {\bibfnamefont {H.-O.}\ \bibnamefont {May}},\ }\bibfield  {title} {\bibinfo {title} {Thermodynamic geometry of the gaussian core model fluid},\ }\href@noop {} {\bibfield  {journal} {\bibinfo  {journal} {Fluid Phase Equilibria}\ ,\ \bibinfo {pages} {113033}} (\bibinfo {year} {2021})}\BibitemShut {NoStop}%
\bibitem [{\citenamefont {Kumara}\ \emph {et~al.}(2022)\citenamefont {Kumara}, \citenamefont {Rizwan}, \citenamefont {Hegde}, \citenamefont {Ali},\ and\ \citenamefont {Ajith}}]{kumara2023microstructure}%
  \BibitemOpen
  \bibfield  {author} {\bibinfo {author} {\bibfnamefont {A.~N.}\ \bibnamefont {Kumara}}, \bibinfo {author} {\bibfnamefont {C.~L.~A.}\ \bibnamefont {Rizwan}}, \bibinfo {author} {\bibfnamefont {K.}~\bibnamefont {Hegde}}, \bibinfo {author} {\bibfnamefont {M.~S.}\ \bibnamefont {Ali}},\ and\ \bibinfo {author} {\bibfnamefont {K.~M.}\ \bibnamefont {Ajith}},\ }\bibfield  {title} {\bibinfo {title} {Microstructure of five-dimensional neutral gauss–bonnet black hole in anti-de sitter spacetime via $p-v$ criticality},\ }\bibfield  {journal} {\bibinfo  {journal} {General Relativity and Gravitation}\ }\textbf {\bibinfo {volume} {55}},\ \href {https://doi.org/10.1007/s10714-022-03050-y} {10.1007/s10714-022-03050-y} (\bibinfo {year} {2022})\BibitemShut {NoStop}%
\bibitem [{\citenamefont {Wei}\ \emph {et~al.}(2019{\natexlab{a}})\citenamefont {Wei}, \citenamefont {Liu},\ and\ \citenamefont {Mann}}]{wei2019repulsive}%
  \BibitemOpen
  \bibfield  {author} {\bibinfo {author} {\bibfnamefont {S.-W.}\ \bibnamefont {Wei}}, \bibinfo {author} {\bibfnamefont {Y.-X.}\ \bibnamefont {Liu}},\ and\ \bibinfo {author} {\bibfnamefont {R.~B.}\ \bibnamefont {Mann}},\ }\bibfield  {title} {\bibinfo {title} {Repulsive interactions and universal properties of charged anti--de sitter black hole microstructures},\ }\href {https://doi.org/10.1103/PhysRevLett.123.071103} {\bibfield  {journal} {\bibinfo  {journal} {Phys. Rev. Lett.}\ }\textbf {\bibinfo {volume} {123}},\ \bibinfo {pages} {071103} (\bibinfo {year} {2019}{\natexlab{a}})}\BibitemShut {NoStop}%
\bibitem [{\citenamefont {Wei}\ \emph {et~al.}(2019{\natexlab{b}})\citenamefont {Wei}, \citenamefont {Liu},\ and\ \citenamefont {Mann}}]{wei2019ruppeiner}%
  \BibitemOpen
  \bibfield  {author} {\bibinfo {author} {\bibfnamefont {S.-W.}\ \bibnamefont {Wei}}, \bibinfo {author} {\bibfnamefont {Y.-X.}\ \bibnamefont {Liu}},\ and\ \bibinfo {author} {\bibfnamefont {R.~B.}\ \bibnamefont {Mann}},\ }\bibfield  {title} {\bibinfo {title} {Ruppeiner geometry, phase transitions, and the microstructure of charged ads black holes},\ }\href {https://doi.org/10.1103/PhysRevD.100.124033} {\bibfield  {journal} {\bibinfo  {journal} {Phys. Rev. D}\ }\textbf {\bibinfo {volume} {100}},\ \bibinfo {pages} {124033} (\bibinfo {year} {2019}{\natexlab{b}})}\BibitemShut {NoStop}%
\bibitem [{\citenamefont {Habicht}\ \emph {et~al.}(2015)\citenamefont {Habicht}, \citenamefont {Schmolke}, \citenamefont {Goerigk}, \citenamefont {Lange}, \citenamefont {Saalwaechter}, \citenamefont {Ballauff},\ and\ \citenamefont {Seiffert}}]{habicht2015critical}%
  \BibitemOpen
  \bibfield  {author} {\bibinfo {author} {\bibfnamefont {A.}~\bibnamefont {Habicht}}, \bibinfo {author} {\bibfnamefont {W.}~\bibnamefont {Schmolke}}, \bibinfo {author} {\bibfnamefont {G.}~\bibnamefont {Goerigk}}, \bibinfo {author} {\bibfnamefont {F.}~\bibnamefont {Lange}}, \bibinfo {author} {\bibfnamefont {K.}~\bibnamefont {Saalwaechter}}, \bibinfo {author} {\bibfnamefont {M.}~\bibnamefont {Ballauff}},\ and\ \bibinfo {author} {\bibfnamefont {S.}~\bibnamefont {Seiffert}},\ }\bibfield  {title} {\bibinfo {title} {Critical fluctuations and static inhomogeneities in polymer gel volume phase transitions},\ }\href@noop {} {\bibfield  {journal} {\bibinfo  {journal} {Journal of Polymer Science Part B: Polymer Physics}\ }\textbf {\bibinfo {volume} {53}},\ \bibinfo {pages} {1112} (\bibinfo {year} {2015})}\BibitemShut {NoStop}%
\bibitem [{\citenamefont {Bishop}\ and\ \citenamefont {Goldberg}(1980)}]{BishopGoldberg1980}%
  \BibitemOpen
  \bibfield  {author} {\bibinfo {author} {\bibfnamefont {R.~L.}\ \bibnamefont {Bishop}}\ and\ \bibinfo {author} {\bibfnamefont {S.~I.}\ \bibnamefont {Goldberg}},\ }\href@noop {} {\emph {\bibinfo {title} {Tensor Analysis on Manifolds}}}\ (\bibinfo  {publisher} {Dover Publications},\ \bibinfo {year} {1980})\BibitemShut {NoStop}%
\bibitem [{\citenamefont {Gray}(1997)}]{Gray1997}%
  \BibitemOpen
  \bibfield  {author} {\bibinfo {author} {\bibfnamefont {A.}~\bibnamefont {Gray}},\ }\bibinfo {title} {Christoffel symbols},\ in\ \href@noop {} {\emph {\bibinfo {booktitle} {Modern Differential Geometry of Curves and Surfaces with Mathematica}}}\ (\bibinfo  {publisher} {CRC Press},\ \bibinfo {address} {Boca Raton, FL},\ \bibinfo {year} {1997})\ pp.\ \bibinfo {pages} {509--513},\ \bibinfo {edition} {2nd}\ ed.\BibitemShut {Stop}%
\bibitem [{\citenamefont {Simo}\ and\ \citenamefont {Miehe}(1992)}]{simo1992associative}%
  \BibitemOpen
  \bibfield  {author} {\bibinfo {author} {\bibfnamefont {J.~C.}\ \bibnamefont {Simo}}\ and\ \bibinfo {author} {\bibfnamefont {C.}~\bibnamefont {Miehe}},\ }\bibfield  {title} {\bibinfo {title} {Associative coupled thermoplasticity at finite strains: Formulation, numerical analysis and implementation},\ }\href@noop {} {\bibfield  {journal} {\bibinfo  {journal} {Computer Methods in Applied Mechanics and Engineering}\ }\textbf {\bibinfo {volume} {98}},\ \bibinfo {pages} {41} (\bibinfo {year} {1992})}\BibitemShut {NoStop}%
\bibitem [{\citenamefont {Hahne}(2005)}]{hahne2005critical}%
  \BibitemOpen
  \bibfield  {author} {\bibinfo {author} {\bibfnamefont {F.~J.~W.}\ \bibnamefont {Hahne}},\ }\href@noop {} {\emph {\bibinfo {title} {Critical Phenomena: Proceedings of the Summer School Held at the University of Stellenbosch, South Africa January 18--29, 1982}}},\ Vol.\ \bibinfo {volume} {186}\ (\bibinfo  {publisher} {Springer},\ \bibinfo {year} {2005})\BibitemShut {NoStop}%
\bibitem [{\citenamefont {Das}\ and\ \citenamefont {Roy}(2021)}]{das2021poroviscoelasticity}%
  \BibitemOpen
  \bibfield  {author} {\bibinfo {author} {\bibfnamefont {S.}~\bibnamefont {Das}}\ and\ \bibinfo {author} {\bibfnamefont {D.}~\bibnamefont {Roy}},\ }\bibfield  {title} {\bibinfo {title} {A poroviscoelasticity model based on effective temperature for water and temperature driven phase transition in hydrogels},\ }\href@noop {} {\bibfield  {journal} {\bibinfo  {journal} {International Journal of Mechanical Sciences}\ }\textbf {\bibinfo {volume} {196}},\ \bibinfo {pages} {106290} (\bibinfo {year} {2021})}\BibitemShut {NoStop}%
\bibitem [{\citenamefont {Tanaka}(1979)}]{TANAKA19791404}%
  \BibitemOpen
  \bibfield  {author} {\bibinfo {author} {\bibfnamefont {T.}~\bibnamefont {Tanaka}},\ }\bibfield  {title} {\bibinfo {title} {Phase transitions in gels and a single polymer},\ }\href {https://doi.org/https://doi.org/10.1016/0032-3861(79)90281-7} {\bibfield  {journal} {\bibinfo  {journal} {Polymer}\ }\textbf {\bibinfo {volume} {20}},\ \bibinfo {pages} {1404} (\bibinfo {year} {1979})}\BibitemShut {NoStop}%
\bibitem [{\citenamefont {Pallas}\ and\ \citenamefont {Pethica}(1985)}]{pallas1985}%
  \BibitemOpen
  \bibfield  {author} {\bibinfo {author} {\bibfnamefont {N.~R.}\ \bibnamefont {Pallas}}\ and\ \bibinfo {author} {\bibfnamefont {B.~A.}\ \bibnamefont {Pethica}},\ }\bibfield  {title} {\bibinfo {title} {Liquid-expanded to liquid-condensed transition in lipid monolayers at the air/water interface},\ }\href {https://doi.org/10.1021/la00064a019} {\bibfield  {journal} {\bibinfo  {journal} {Langmuir}\ }\textbf {\bibinfo {volume} {1}},\ \bibinfo {pages} {509} (\bibinfo {year} {1985})},\ \Eprint {https://arxiv.org/abs/https://doi.org/10.1021/la00064a019} {https://doi.org/10.1021/la00064a019} \BibitemShut {NoStop}%
\bibitem [{\citenamefont {Hifeda}\ and\ \citenamefont {Rayfield}(1985)}]{HIFEDA1985}%
  \BibitemOpen
  \bibfield  {author} {\bibinfo {author} {\bibfnamefont {Y.}~\bibnamefont {Hifeda}}\ and\ \bibinfo {author} {\bibfnamefont {G.}~\bibnamefont {Rayfield}},\ }\bibfield  {title} {\bibinfo {title} {Phase transitions in fatty acid monolayers containing a single double bond in the fatty acid tail},\ }\href {https://doi.org/https://doi.org/10.1016/0021-9797(85)90025-6} {\bibfield  {journal} {\bibinfo  {journal} {Journal of Colloid and Interface Science}\ }\textbf {\bibinfo {volume} {104}},\ \bibinfo {pages} {209} (\bibinfo {year} {1985})}\BibitemShut {NoStop}%
\bibitem [{\citenamefont {Hifeda}\ and\ \citenamefont {Rayfield}(1992)}]{Hifeda1992}%
  \BibitemOpen
  \bibfield  {author} {\bibinfo {author} {\bibfnamefont {Y.}~\bibnamefont {Hifeda}}\ and\ \bibinfo {author} {\bibfnamefont {G.}~\bibnamefont {Rayfield}},\ }\bibfield  {title} {\bibinfo {title} {Evidence for first-order phase transitions in lipid and fatty acid monolayers},\ }\href {https://doi.org/10.1021/la00037a036} {\bibfield  {journal} {\bibinfo  {journal} {Langmuir}\ }\textbf {\bibinfo {volume} {8}},\ \bibinfo {pages} {197 – 200} (\bibinfo {year} {1992})}\BibitemShut {NoStop}%
\bibitem [{\citenamefont {Crane}\ \emph {et~al.}(1999)\citenamefont {Crane}, \citenamefont {Putz},\ and\ \citenamefont {Hall}}]{CRANE1999}%
  \BibitemOpen
  \bibfield  {author} {\bibinfo {author} {\bibfnamefont {J.~M.}\ \bibnamefont {Crane}}, \bibinfo {author} {\bibfnamefont {G.}~\bibnamefont {Putz}},\ and\ \bibinfo {author} {\bibfnamefont {S.~B.}\ \bibnamefont {Hall}},\ }\bibfield  {title} {\bibinfo {title} {Persistence of phase coexistence in disaturated phosphatidylcholine monolayers at high surface pressures},\ }\href {https://doi.org/https://doi.org/10.1016/S0006-3495(99)77143-2} {\bibfield  {journal} {\bibinfo  {journal} {Biophysical Journal}\ }\textbf {\bibinfo {volume} {77}},\ \bibinfo {pages} {3134} (\bibinfo {year} {1999})}\BibitemShut {NoStop}%
\bibitem [{\citenamefont {Tanaka}(1996)}]{tanaka1996coarsening}%
  \BibitemOpen
  \bibfield  {author} {\bibinfo {author} {\bibfnamefont {H.}~\bibnamefont {Tanaka}},\ }\bibfield  {title} {\bibinfo {title} {Coarsening mechanisms of droplet spinodal decomposition in binary fluid mixtures},\ }\href@noop {} {\bibfield  {journal} {\bibinfo  {journal} {The Journal of chemical physics}\ }\textbf {\bibinfo {volume} {105}},\ \bibinfo {pages} {10099} (\bibinfo {year} {1996})}\BibitemShut {NoStop}%
\bibitem [{\citenamefont {Clerk-Maxwell}(1875)}]{clerk1875dynamical}%
  \BibitemOpen
  \bibfield  {author} {\bibinfo {author} {\bibfnamefont {J.}~\bibnamefont {Clerk-Maxwell}},\ }\bibfield  {title} {\bibinfo {title} {On the dynamical evidence of the molecular constitution of bodies},\ }\href@noop {} {\bibfield  {journal} {\bibinfo  {journal} {Nature}\ }\textbf {\bibinfo {volume} {11}},\ \bibinfo {pages} {357} (\bibinfo {year} {1875})}\BibitemShut {NoStop}%
\bibitem [{\citenamefont {Kumar}\ and\ \citenamefont {Sarkar}(2014)}]{sarkar2014}%
  \BibitemOpen
  \bibfield  {author} {\bibinfo {author} {\bibfnamefont {P.}~\bibnamefont {Kumar}}\ and\ \bibinfo {author} {\bibfnamefont {T.}~\bibnamefont {Sarkar}},\ }\bibfield  {title} {\bibinfo {title} {Geometric critical exponents in classical and quantum phase transitions},\ }\href {https://doi.org/10.1103/PhysRevE.90.042145} {\bibfield  {journal} {\bibinfo  {journal} {Phys. Rev. E}\ }\textbf {\bibinfo {volume} {90}},\ \bibinfo {pages} {042145} (\bibinfo {year} {2014})}\BibitemShut {NoStop}%
\bibitem [{\citenamefont {Bolmatov}\ \emph {et~al.}(2013)\citenamefont {Bolmatov}, \citenamefont {Brazhkin},\ and\ \citenamefont {Trachenko}}]{bolmatov2013thermodynamic}%
  \BibitemOpen
  \bibfield  {author} {\bibinfo {author} {\bibfnamefont {D.}~\bibnamefont {Bolmatov}}, \bibinfo {author} {\bibfnamefont {V.}~\bibnamefont {Brazhkin}},\ and\ \bibinfo {author} {\bibfnamefont {K.}~\bibnamefont {Trachenko}},\ }\bibfield  {title} {\bibinfo {title} {Thermodynamic behaviour of supercritical matter},\ }\href@noop {} {\bibfield  {journal} {\bibinfo  {journal} {Nature communications}\ }\textbf {\bibinfo {volume} {4}},\ \bibinfo {pages} {2331} (\bibinfo {year} {2013})}\BibitemShut {NoStop}%
\bibitem [{\citenamefont {Touchette}(2009)}]{TOUCHETTE20091}%
  \BibitemOpen
  \bibfield  {author} {\bibinfo {author} {\bibfnamefont {H.}~\bibnamefont {Touchette}},\ }\bibfield  {title} {\bibinfo {title} {The large deviation approach to statistical mechanics},\ }\href {https://doi.org/https://doi.org/10.1016/j.physrep.2009.05.002} {\bibfield  {journal} {\bibinfo  {journal} {Physics Reports}\ }\textbf {\bibinfo {volume} {478}},\ \bibinfo {pages} {1} (\bibinfo {year} {2009})}\BibitemShut {NoStop}%
\end{thebibliography}
\end{document}